\documentclass[acmlarge]{acmart}

\usepackage{color}
\usepackage{subcaption}
\usepackage{bbm}
\usepackage{bm}
\usepackage[ruled,vlined]{algorithm2e}
\SetKwRepeat{Do}{do}{while}

\newcommand{\norm}[1]{\left\lVert#1\right\rVert}

\newcommand{\menge}[1]{\left\lbrace#1\right\rbrace}

\newcommand{\intD}{\;\mathrm{d}}
\newcommand{\rom}[1]{\uppercase\expandafter{\romannumeral #1\relax}}

\newcommand{\inner}[1]{\left< #1 \right>}

\AtBeginDocument{%
    \providecommand\BibTeX{{%
    \normalfont B\kern-0.5em{\scshape i\kern-0.25em b}\kern-0.8em\TeX}}}

\setcopyright{acmcopyright}
\copyrightyear{2022}
\acmYear{2022}

\acmJournal{TOMS}

\author{Jonas Kusch}
\affiliation{%
  \institution{University of Innsbruck}
  \streetaddress{Technikerstraße 13}
  \city{Innsbruck}
  \country{Austria}}
\email{jonas.kusch1@gmail.com}

\author{Steffen Schotth{\"o}fer}
\affiliation{%
  \institution{Karlsruhe Institute of Technology}
  \streetaddress{Englerstraße 2}
  \city{Karlsruhe}
  \country{Germany}}
\email{steffen.schotthoefer@kit.edu}

\author{Pia Stammer}
\affiliation{%
  \institution{Karlsruhe Institute of Technology}
  \streetaddress{Englerstraße 2}
  \city{Karlsruhe}
  \country{Germany}}
 \affiliation{%
  \institution{German Cancer Research Center - DKFZ}
  \city{Heidelberg}
  \country{Germany}}
  \affiliation{%
  \institution{HIDSS4Health - Helmholtz Information and Data Science School for Health}
  \city{Karlsruhe/Heidelberg}
  \country{Germany}}
\email{pia.stammer@kit.edu}

\author{Jannick Wolters}
\affiliation{%
  \institution{Karlsruhe Institute of Technology}
  \streetaddress{Englerstraße 2}
  \city{Karlsruhe}
  \country{Germany}}
\email{jannick.wolters@rwth-aachen.de}

\author{Tianbai Xiao}
\affiliation{%
  \institution{Karlsruhe Institute of Technology}
  \streetaddress{Englerstraße 2}
  \city{Karlsruhe}
  \country{Germany}}
\email{tianbai.xiao@kit.edu}

\begin{document}

\title{KiT-RT: An extendable framework for radiative transfer and therapy}

\begin{abstract}
    In this paper we present KiT-RT (Kinetic Transport Solver for Radiation Therapy), an open-source C++ based framework for solving kinetic equations in radiation therapy applications. The aim of this code framework is to provide a collection of classical deterministic solvers for unstructured meshes that allow for easy extendability. Therefore, KiT-RT is a convenient base to test new numerical methods in various applications and compare them against conventional solvers. The implementation includes spherical-harmonics, minimal entropy, neural minimal entropy and discrete ordinates methods. Solution characteristics and efficiency are presented through several test cases ranging from radiation transport to electron radiation therapy. Due to the variety of included numerical methods and easy extendability, the presented open source code is attractive for both developers, who want a basis to build their own numerical solvers and users or application engineers, who want to gain experimental insights without directly interfering with the codebase.
\end{abstract}


\maketitle

\section{Introduction}\label{sec:intro}

Personalized medicine in radiation oncology has been an important research topic in the last decades. To allow for accurate, reliable and efficient treatment planning tailored towards individual patient needs, there is a growing desire to undertake direct numerical simulation for radiation therapy. High-fidelity numerical solutions enable quantitative estimation of the dose received by the tumor as well as the surrounding tissue, while allowing for an automated generation of optimal treatment plans. Besides the aim to ensure sufficient accuracy, such simulations are required to run on limited computational resources such as workstation PCs. 

Traditional methods to predict dose distributions in radiation oncology largely rely on simplified models, such as pencil beam models based on the Fermi-Eyges theory \cite{eyges1948multiple}. While such models are computationally efficient, they often lack the required accuracy, especially in cases including air cavities or other inhomogeneities \cite{hogstrom1981electron, krieger2005montecarlo}. On the other hand, Monte Carlo (MC) algorithms, which simulate individual interacting particles, achieve a satisfactory accuracy \cite{andreo1991montecarlo}. However, despite ongoing research to accelerate MC methods, their high computational costs currently renders them impractical for clinical usage \cite{fippel2004monte, jia2012gpu}. To obtain a computationally feasible model with comparable accuracy, radiation particles are described on a mesoscopic level through the deterministic linear Boltzmann equation \cite{borgers1998complexity,tervo1999finite,tervo2002inverse,tervo2008optimal}. An efficient and accurate numerical approximation to the linear Boltzmann equation can be achieved through the construction of grid-based macroscopic approximations \cite{duclous2010deterministic, vassiliev2010validation, vassiliev2008feasibility, gifford2006comparison}. Variants of grid based methods for radiation therapy can be found in for example \cite{hensel2006deterministic, barnard2012optimal, huang2015application, olbrant2010generalized,kuepper2016models,kusch2021robust}.

The available grid-based methods employ various macroscopic approximations to the linear Boltzmann equation which all exhibit certain advantages and shortcomings. 

In \cite{duclous2010deterministic}, the modal entropy method called $M_N$ is used as a macroscopic model. This method preserves positivity of particle distributions while yielding accurate results with little spurious oscillations. However, the need to solve a possibly ill-posed optimization problem in every spatial cell and time step results in increased computational costs. While analytic solutions to the optimization problems are available at small truncation order, they cannot capture all physical effects accurately. Further approaches aim at mitigating the challenge of the problems ill-condition by regularization ~\cite{alldredge2018regularized} or reducing the associated computational costs, e.g. using  neural networks ~\cite{schotthoefer2021structurepreserving}.

In \cite{olbrant2010generalized} the use of computationally cheap nodal discretizations, known as the $S_N$ method has been proposed. In this case, the solution remains positive and it is shown that the expensive scattering terms can be approximated efficiently with a Fokker-Planck approximation. Furthermore, the solution can exhibit non-physical artifacts, known as ray-effects \cite{lathrop1968ray,mathews1999propagation,morel2003analysis}, which reduce the approximation quality. While methods to mitigate ray-effects exist, see e.g.~\cite{Camminady2019RayEM,frank2020ray,abu2001angular,lathrop1971remedies,tencer2016ray}, they commonly require picking problem-dependent tuning parameters.

In \cite{kuepper2016models}, the modal $P_N$ method has been employed to derive a macroscopic model for radiation treatment planning. While it does not preserve positivity of the solution and can potentially yield oscillatory approximations, it allows for an efficient numerical treatment of scattering terms. In \cite{kusch2021robust} a combination of $S_N$ and $P_N$ methods which reduces computational costs through a dynamical low rank approximation \cite{KochLubich07} has been proposed. The efficiency of this method relies on the ability to describe the solution through a low-rank function.

The variety of different methods allows for individual method choices tailored to different settings. The comparability of different methods in a uniform framework is not only interesting for clinical usage, but also for future research in computational radiation therapy. Our aim of the open-source C++ \textit{Kinetic Transport for radiation therapy} (KiT-RT) framework is therefore to provide a collection of available deterministic methods. Special focus is put on extendability by the use of polymorphism to simplify the implementation of novel solution methods. The methods provided by our framework are optimized for an application on workstation PCs. This meets the typical requirements in radiation therapy applications: For clinical usage the computational resources are often limited and the time between recording the CT image and the actual treatment must not exceed a certain time period. Hence, radiation therapy codes that are applicable for clinical use should run efficiently on workstation PCs. Moreover, conventional codes often require structured grids \cite{seibold2014starmap,wieser2017development, perl2012topas,GarretHauck,KRISTOPHERGARRETT2015573}, leading to inaccurate representations of structures on CT images. While accuracy in practice is also limited by the CT density values which are given on a structured grid, these are often downsampled to a lower resolution for dose computations. Further, a recomputation of the CT values for unstructured grids is feasible if it improves the quality of dose computations. Therefore, our framework provides functionalities for both unstructured meshes which preserve organ outlines on CT images, as well as standard rectangular grids.

This paper aims at presenting the developed framework and its functionality while providing the necessary mathematical and physical background on principles the software is based on. In Section~\ref{sec:theory} we provide the underlying physical model as well as its reformulation as a time-dependent partial differential equation. Section~\ref{sec:macroModel} discusses different macroscopic models as well as advantages and disadvantages of the individual underlying directional discretizations. Sections~\ref{sec:discretization} and~\ref{sec:softwareArch} focus on the used discretizations and software architecture, respectively. Lastly, we validate our implementations and analyse their performance for different test cases in Section~\ref{sec:validation}.

\section{Physical model}\label{sec:theory}
Let us recap the main physical model that is used for computational radiotherapy treatment planning. The main goal is to compute the radiation dose distribution
\begin{equation}
    D(\mathbf x)=\frac{1}{\rho(\mathbf x)}\int_0^{\infty}\int_{\mathbb{S}^2}S(E,\mathbf x)\psi(E,\mathbf x,\mathbf \Omega)\intD\mathbf \Omega \intD E\;,
\end{equation}
that results from a given treatment plan. Here, $E\in\mathbb{R}_+$ is the energy, $\mathbf{x}\in \mathbf{X}\subset\mathbb{R}^3$ denotes the spatial domain, and $\mathbf{\Omega}\in\mathbb{S}^2$ is the flight direction of particles. Moreover, $\psi:\mathbb{R}_+\times\mathbb{R}^3\times\mathbb{S}^2\rightarrow\mathbb{R}$ denotes the radiation flux density and $\rho:\mathbb{R}^3\rightarrow\mathbb{R}$ is the patient tissue density. The stopping power $S:\mathbb{R}_+\times\mathbb{R}^3 \rightarrow \mathbb{R}$ models the continuous energy loss of particles due to scattering with tissue and is defined as
\begin{equation}
    S(E,\mathbf x) = \int_0^{\infty} E'\int_{-1}^1\Sigma(E,E',\mathbf x,\mu)\intD\mu \intD E'
\end{equation}
with the scattering cross section $\Sigma:\mathbb{R}_+\times \mathbb{R}_+\times \mathbb{R}^3\times[-1,1]\rightarrow \mathbb{R}$. The radiation flux density, which describes the probability of finding a particle at a certain region in phase space, can be computed from the continuous slowing down (CSD) approximation \cite{larsen1997electron} of the energy dependent linear Boltzmann equation
\begin{equation}\label{eq:CSDorig}
\begin{aligned}
    &-\partial_E\left(S(E,\mathbf x)\psi(E,\mathbf x,\mathbf \Omega)\right)+\mathbf \Omega\cdot\nabla_x\psi(E,\mathbf x,\mathbf \Omega)+\Sigma_t(E,\mathbf x)\psi(E,\mathbf x,\mathbf \Omega) \\
    &= \int_{\mathbb{S}^2}\Sigma_s(E,\mathbf x,\mathbf \Omega\cdot\mathbf \Omega')\psi(E,\mathbf x,\mathbf \Omega')\intD\mathbf \Omega'\;.
\end{aligned}
\end{equation}
This model assumes a continuous energy loss of particles traveling through a background material, which is modeled using the stopping power $S$. The scattering cross section $\Sigma_s(E,\mathbf x,\mathbf \Omega\cdot\mathbf \Omega')$ denotes the probability of particles at position $\mathbf x$ with energy $E$ changing their flight direction from $\mathbf \Omega'$ to $\mathbf\Omega$ due to a collision with the patient tissue. The total cross section $\Sigma_t$ is given by
\begin{equation}
    \Sigma_t(E,\mathbf x) = \Sigma_{s,0}(E,\mathbf x)=2\pi \int_{-1}^1\Sigma_s(E,\mathbf x,\mu)\intD \mu\;.
\end{equation}
To simplify the evaluation of material properties, we follow the common assumption that all materials are water-equivalent and differ only in density \cite[e.g.][]{woo1990validity, olbrant2010generalized,kuepper2016models}, i.e.,
\begin{equation}
\begin{aligned}
    &S(E,\mathbf x) = S^{H_2O}(E)\rho(\mathbf x), \\
    &\Sigma_t(E,\mathbf x) = \Sigma_t^{H_2O}(E)\rho(\mathbf x), \\
    &\Sigma_s(E,\mathbf x,\mathbf \Omega\cdot\mathbf \Omega') = \Sigma_s^{H_2O}(E,\mathbf \Omega\cdot\mathbf \Omega')\rho(\mathbf x)\;,
\end{aligned}
\end{equation}
where we leave out the superscript $H_2O$ in the following. Cross-sections for water are taken from the ICRU database \cite{gregoire2011state}.  Having defined the prerequisites of our physical model, we can focus on bringing it into a form that allows for computing numerical approximations efficiently. It turns out that the energy variable in \eqref{eq:CSDorig} can be treated as a pseudo-time, which facilitates solving the CSD equation. For a given maximal energy $E_{\mathrm{max}}$ let us define the transformed energy as
\begin{equation}
    \widetilde{E}(E) := \int_0^{E_{\mathrm{max}}} \frac{1}{S(E')}\intD E'-\int_0^E \frac{1}{S(E')}\intD E'
\end{equation}
and the transformed particle density as
\begin{align}
    \widetilde{\psi}(\widetilde E,\mathbf x,\mathbf \Omega):= S(E)\rho(\mathbf x)\psi(E(\widetilde E),\mathbf x,\mathbf \Omega)\;.
\end{align}
Then, multiplying \eqref{eq:CSDorig} with $S(E)$ and plugging in the defined transformation gives
\begin{equation}
    \partial_{\widetilde{E}}\widetilde{\psi}(\widetilde{E},\mathbf x,\mathbf \Omega)+\mathbf \Omega\cdot\nabla_x \frac{\widetilde{\psi}(\widetilde{E},\mathbf x,\mathbf \Omega)}{\rho(\mathbf{x})}+\widetilde\Sigma_t(\widetilde E)\widetilde{\psi}(\widetilde{E},\mathbf x,\mathbf \Omega) = \int_{\mathbb{S}^2}\widetilde\Sigma_s(\widetilde{E},\mathbf \Omega\cdot\mathbf \Omega')\widetilde{\psi}(\widetilde{E},\mathbf x,\mathbf \Omega')\intD \mathbf \Omega'\;,
\label{CSD6}
\end{equation}
where we define $\widetilde\Sigma_{t}(\widetilde E):=\Sigma_t(E(\widetilde E))$ and $\widetilde\Sigma_{s}(\widetilde E,\mathbf \Omega\cdot\mathbf \Omega'):=\Sigma_s(E(\widetilde E),\mathbf \Omega\cdot\mathbf \Omega')$. Dropping the tilde notation and treating $\widetilde E$ as a pseudo-time $t$ gives a slightly modified version of the classical linear Boltzmann equation
\begin{equation}\label{eq:BoltzmannCSDTrafo}
\begin{aligned}
    \partial_{t}\psi(t,\mathbf x,\mathbf \Omega)+&\mathbf \Omega\cdot\nabla_x \frac{\psi(t,\mathbf x,\mathbf \Omega)}{\rho(\mathbf{x})}+\Sigma_t(t)\psi(t,\mathbf x,\mathbf \Omega) = \int_{\mathbb{S}^2}\Sigma_s(t,\mathbf \Omega\cdot\mathbf \Omega')\psi(t,\mathbf x,\mathbf \Omega')\intD \mathbf \Omega'\\
    &\psi(t=0,\mathbf x,\mathbf \Omega) = S(E_{\text{max}})\rho(\mathbf x)\psi(E_{\text{max}},\mathbf x,\mathbf \Omega)\;.
\end{aligned}
\end{equation}
Hence, the CSD equation can be treated numerically with classical closure methods and space-time discretizations. Let us first discuss methods to discretize the directional domain, yielding macroscopic evolution equations.

\section{Macroscopic models}\label{sec:macroModel}
This section discusses macroscopic models to \eqref{eq:BoltzmannCSDTrafo}. These models are derived from nodal and modal discretizations of the directional domain. 
\subsection{Modal discretizations}
Modal discretizations of \eqref{eq:BoltzmannCSDTrafo} can be interpreted as a closure problem~\cite{Levermore,Levermore1996MomentCH}. To present the derivation of different closures, we first formulate the moment equations which are not closed. Second, we close these equations with the $P_N$ closure and third, we derive the $M_N$ closure.
\subsubsection*{Moment equations}
Let us derive an evolution equation to describe the moments of the radiation flux with respect to the real-valued spherical harmonics basis functions. These are defined as the real parts of the spherical harmonics
\begin{align*}
Y_{\ell}^k(\mathbf{\Omega}) = \sqrt{\frac{2\ell +1}{4\pi}\frac{(\ell-k)!}{(\ell+k)!}}\ e^{ik\varphi}P_{\ell}^k(\mu)\;,
\end{align*}
where $P_{\ell}^k$ are the associated Legendre polynomials. Then, the real spherical harmonics are given as
\begin{align*}
    m_{\ell}^k(\mathbf{\Omega}) = 
    \begin{cases}
        \frac{(-1)^k}{\sqrt{2}}\left( Y_{\ell}^k(\mathbf{\Omega}) + (-1)^k Y_{\ell}^{-k}(\mathbf{\Omega}) \right), & k > 0\;, \\
        Y_{\ell}^0(\mathbf{\Omega}) & k = 0 \;, \\
        -\frac{(-1)^k i}{\sqrt{2}}\left( Y_{\ell}^{-k}(\mathbf{\Omega}) - (-1)^k Y_{\ell}^{k}(\mathbf{\Omega}) \right), & k < 0\;,
    \end{cases}
\end{align*}
where $i$ is the imaginary unit. Collecting all basis functions up to degree $N$ in a vector
\begin{align*}
    \mathbf m = \left(m_0^0, m_1^{-1}, m_1^{0}, m_1^{1},\cdots, m_N^{N}\right)^T\in\mathbb{R}^{(N+1)^2}
\end{align*}
yields the so-called moments
\begin{align*}
    u_{\ell}^k(t,\mathbf x) := \int_{\mathbb{S}^2} \psi(t,\mathbf x,\mathbf\Omega) m_{\ell}^k(\mathbf \Omega) \intD \mathbf\Omega\;.
\end{align*}
Evolution equations for the moments can be derived by testing \eqref{eq:BoltzmannCSDTrafo} against $\mathbf{m}_{\ell} = (m_{\ell}^{-\ell},\cdots,m_{\ell}^{\ell})$, which gives
\begin{align}
    \partial_{t}\mathbf u_{\ell}(t,\mathbf x)+&\nabla_x\cdot\int_{ \mathbb{S}^2}\mathbf\Omega\mathbf m_{\ell}(\mathbf\Omega)\frac{\psi(t,\mathbf x,\mathbf\Omega)}{\rho(\mathbf x)}\intD \mathbf{\Omega}+\Sigma_t(t)\mathbf u_{\ell}(t,\mathbf x)\nonumber\\
    &= \int_{\mathbb{S}^2}\int_{\mathbb{S}^2}\mathbf m_{\ell}(\mathbf\Omega)\Sigma_s(t,\mathbf\Omega\cdot\mathbf\Omega')\psi(t,\mathbf x,\mathbf\Omega')\intD \mathbf\Omega'\intD \mathbf\Omega\;.
\end{align}
To rewrite this equation, we use the spherical harmonics recursion relation
\begin{align*}
    \Omega_i \mathbf{m}_{\ell} = \mathbf{a}_{\ell}^i\mathbf m_{\ell-1} + \mathbf{a}_{\ell+1}^i\mathbf m_{\ell+1} \enskip \text{ with } \mathbf{a}_{\ell}^i\in\mathbb{R}^{(2\ell-1)\times (2\ell+1)}
\end{align*}
as well as the fact that there exists a diagonal matrix $\bm{\Sigma}_{\ell}(t)$ with entries $\Sigma_{\ell,kk} = \Sigma_{\ell}^k := 2\pi\int_{[-1,1]}P_{\ell}(\mu)\Sigma_s(t,\mu)\intD \mu$ such that
\begin{align*}
    \int_{\mathbb{S}^2}\int_{\mathbb{S}^2}\mathbf m_{\ell}(\mathbf\Omega)\Sigma_s(t,\mathbf\Omega\cdot\mathbf\Omega')\psi(t,\mathbf x,\mathbf\Omega')\intD \mathbf\Omega'd\mathbf\Omega = \bm{\Sigma}_{\ell}(t) \mathbf u_{\ell}(t,\mathbf x)\;.
\end{align*}
Then, the moment equations at degree $\ell$ become
\begin{align}
    \partial_{t}\mathbf u_{\ell}(t,\mathbf x)+&\sum_{i=1}^3\partial_{x_i}\left(\mathbf{a}_{\ell}^i\mathbf u_{\ell-1}(t,\mathbf x) + \mathbf{a}_{\ell+1}^i\mathbf u_{\ell+1}(t,\mathbf x)\right)/\rho(\mathbf{x})+\Sigma_t(t)\mathbf u_{\ell}(t,\mathbf x)\nonumber\\
    &= \bm{\Sigma}_{\ell}(t) \mathbf u_{\ell}(t,\mathbf x)\;.
\end{align}
Note that the equations for degree $\ell$ depend on the moments of degree $\ell+1$. Hence, to obtain a closed system of moments up to a fixed degree $N$, we need to define a closure relation 
\begin{align}
    \mathbf u_{N+1}(t,\mathbf x)\simeq \mathcal{U}(\mathbf u_{0}(t,\mathbf x),\cdots,\mathbf u_{N}(t,\mathbf x))\;.
\end{align}

\subsubsection*{$P_N$ closure}

The most commonly used closure is the $P_N$ closure which leads to the spherical harmonics ($P_N$) method \cite{case1967linear}. It expands the solution by spherical harmonics, i.e.,
\begin{align}
    \psi(t,\mathbf{x},\mathbf{\Omega}) \approx \psi_{\mathrm{P}_N}(t,\mathbf{x},\mathbf{\Omega}) := \mathbf{u}(t,\mathbf x)^T\mathbf{m}(\mathbf{\Omega})\;,
\end{align}
where $\mathbf{u}\in\mathbb{R}^{(N+1)^2}$ collects all moments according to $\mathbf u = \left(u_0^0, u_1^{-1}, u_1^{0}, u_1^{1},\cdots, u_N^{N}\right)^T\in\mathbb{R}^{(N+1)^2}$.
Hence, the $P_N$ closure is simply given as $\mathcal{U}_{\mathrm{P}_N}\equiv \mathbf 0$. In this case, the moment equations read
\begin{align*}
    \partial_t \mathbf u (t,\mathbf x) =-\mathbf A\cdot\nabla_{\mathbf{x}} \frac{\mathbf u(t,\mathbf x)}{\rho(\mathbf x)}-\Sigma_t(t)\mathbf u (t,\mathbf x)+\bm{\Sigma}\mathbf u (t,\mathbf x),\;
\end{align*}
where $\mathbf A\cdot\nabla_{\mathbf{x}} := \mathbf A_1\partial_{x} + \mathbf A_2\partial_y+ \mathbf A_3\partial_z$ with $\mathbf A_i := \int_{\mathbb{S}^2}\mathbf m\mathbf m^T \Omega_i \intD \mathbf{\Omega}$ and $\bm \Sigma = \mathrm{diag}\left(\Sigma_0^0, \Sigma_1^{-1}, \Sigma_1^{0}, \Sigma_1^{1},\cdots, \Sigma_N^{N}\right)$. While $P_N$ is a computationally efficient method (especially for scattering terms), it does not preserve positivity of the radiation flux approximation and can lead to spurious oscillations. A closure which mitigates oscillations and preserves positivity at significantly increased computational costs is the $M_N$ closure.
\\
\subsubsection*{$M_N$ closure}

The $M_N$ closure~\cite{Levermore1996MomentCH,Levermore} employs the principle of minimal mathematical, i.e., maximal physical entropy to close the moment system.
To this end, we define the twice differentiable, strictly convex kinetic entropy density $\eta:\mathbb{R}_+\rightarrow\mathbb{R}$. Different, problem specific entropy densities can be defined, e.g. the Maxwell-Boltzmann entropy $\eta(g)=g\log(g)-g$, or a quadratic entropy $\eta(g)=g^2$, which recovers the $P_N$ method.
Thus, one can close the system by choosing the reconstructed radiation flux density $\psi_{\mathbf u}$ out of the set of all possible functions 
\begin{align}
    F_{\mathbf m}=\menge{g(t,x,\mathbf{\Omega})>0 : u=\int_{\mathbb{S}^2}{\mathbf m g}\intD \mathbf{\Omega}<\infty},
\end{align}
that fulfill the moment condition $\mathbf u(t,\mathbf{x})=\inner{\mathbf m g}$ as the one with minimal entropy. The modal basis $\mathbf m$ can be chosen arbitrarily. Common choices consist of spherical harmonics or other polynomial basis functions. The minimal entropy closure can be formulated as a constrained optimization problem for a given vector of moments $\mathbf u$,
\begin{align}\label{eq_entropyOCP} 
\min_{g\in F_{\mathbf m}} \int_{\mathbb{S}^2}\eta(g)\intD \mathbf{\Omega} \quad  \text{ s.t. }  \mathbf u=\int_{\mathbb{S}^2}{\mathbf m g}\intD \mathbf{\Omega}.
\end{align}
The minimal value of the objective function is denoted by $h(u)=\inner{\eta(\psi_{\mathbf u})}$ and describes the systems minimal entropy depending on time and space. $\psi_{\mathbf u}$ is the minimizer of Eq.~\eqref{eq_entropyOCP}, which we use to close the moment system
\begin{align}
    \partial_{t}\mathbf u_{\ell}(t,\mathbf x)+&\nabla_x\cdot\int_{ \mathbb{S}^2}\mathbf\Omega\mathbf m_{\ell}(\mathbf\Omega)\frac{\psi_u(t,\mathbf x,\mathbf\Omega)}{\rho(\mathbf x)}\intD \mathbf{\Omega}+\Sigma_t(t)\mathbf u_{\ell}(t,\mathbf x)= \Sigma_{\ell}\mathbf u_{\ell} (t,\mathbf x);.
\end{align}
The minimal entropy method preserves important properties of the underlying equation~\cite{Levermore,alldredge2018regularized}, i.e., positivity of the solution, hyperbolicity of the moment system, dissipation of mathematical entropy and the H-Theorem. The minimal entropy closed moment system is invariant under Galilean transforms. Lastly, if collision invariants of the Boltzmann equations are used as modal basis functions, then the moment system yields a local conservation law. \\
The set of all moments corresponding to a radiation flux density $\psi_{\mathbf u}>0$ is called the realizable set 
\begin{align}
    \mathcal{R}=\menge{\mathbf u: \int_{\mathbb{S}^2}{\mathbf m g}\intD \mathbf{\Omega}=\mathbf u,\, g\in F_{\mathbf m}}.
\end{align}
Outside $\mathcal{R}$ the minimal entropy closure problem has no solution.  At the boundary $\partial \mathcal{R}$, the optimization problem becomes singular and $\psi_{\mathbf u}$ consists of a linear combination of dirac distributions. Near $\partial \mathcal{R}$ the entropy closure becomes ill conditioned and thus, a numerical optimizer requires a large amount of iterations to compute a solution.

To mitigate this issue, a regularized version of the entropy closure problem has been proposed by~\cite{alldredge2018regularized},
\begin{align}\label{eq_entropyOCP_reg} 
\inf_{g\in F_{\mathbf m}}  \int_{\mathbb{S}^2}\eta(g)\intD \mathbf{\Omega}+
\frac{1}{2\gamma}\norm{ \int_{\mathbb{S}^2}{\mathbf m g}\intD \mathbf{\Omega} - \mathbf u}^2_2,
\end{align}
where $\gamma$ is the regularization parameter. Generally, moments of the regularized reconstructed radiation flux density $\int_{\mathbb{S}^2}\mathbf m\psi_{\mathbf u}\intD \mathbf{\Omega}$ deviate from the non-regularized moments. 
For $\gamma\rightarrow 0$, we recover the original entropy closure of Eq.~\eqref{eq_entropyOCP} and the moments coincide again. The regularized entropy closure is solvable for any $\mathbf u\in\mathbb{R}^{(N+1)^2}$ and preserves all structural properties of the non-regularized entropy closure~\cite{alldredge2018regularized}. One can also choose to regularize only parts of the entropy closure, e.g. to preserve moments of specific interest. Then the partially regularized entropy closure reads
\begin{align}\label{eq_entropyOCP_part_reg} 
\inf_{g\in F_m}  \int_{\mathbb{S}^2}\eta(g)\intD \mathbf{\Omega} + \frac{1}{2\gamma}\norm{\int_{\mathbb{S}^2}{\mathbf m^r g} \intD \mathbf{\Omega} - u^r}^2_2\quad  \text{ s.t. }  \mathbf u^{nr}=\int_{\mathbb{S}^2}{\mathbf m^{nr}g}\intD \mathbf{\Omega},
\end{align}
where $\mathbf u^{nr}$ denotes non-regularized moment elements and $\mathbf u^{r}$ denotes regularized elements of the moment vector.\\
Recently, structure preserving deep neural networks have been successfully employed to approximate the entropy closure~\cite{schotthoefer2021structurepreserving} to accelerate the $M_N$ method. The authors leverage convexity of the optimization problem and use corresponding input convex neural networks~\cite{amosICNN} to preserve structural properties of the closure in the neural network based approximation.

\subsection{Nodal discretizations}
The $S_N$ method \cite{lewis1984computational} employs a nodal discretization for the directional domain. To facilitate the computation of integral terms that arise due to scattering, the nodal point sets are commonly chosen according to a quadrature rule.
In the application case of radiative transport, the directional domain is assumed to be the unit sphere $\mathbb{S}^2\subset\mathbb{R}^3$, thus a suitable parametrization is given by spherical coordinates
\begin{align}
    \mathbb{S}^2 =  \menge{ \begin{pmatrix}
           \sqrt{1-\mu^2}\sin(\varphi) \\
           \sqrt{1-\mu^2}\cos(\varphi) \\
           \mu
         \end{pmatrix}
     : \mu\in\left[-1,1\right],\, \varphi\in\left[0,2\pi\right)}.
\end{align}
Note, that we can allow different particle velocities by scaling the unit sphere with a given maximum velocity.
The implementation assumes a slab geometry setting, i.e., lower dimensional velocity spaces are described by a projection of $\mathbb{S}^2$ onto $\mathbb{R}^2$ and $\mathbb{R}$, respectively. Thus, the parametrization of the two-dimensional slab geometry is given by
\begin{align}
    P_{\mathbb{R}^2}(\mathbb{S}^2) =  \menge{ \begin{pmatrix}
           \sqrt{1-\mu^2}\sin(\varphi) \\
           \sqrt{1-\mu^2}\cos(\varphi)
         \end{pmatrix}
     : \mu\in\left[-1,1\right],\, \varphi\in\left[0,2\pi\right)}
\end{align}
and the one dimensional case is described by
\begin{align}
    P_{\mathbb{R}}(\mathbb{S}^2) =  \menge{ \mu     : \mu\in\left[-1,1\right]}
\end{align}
Hence the task is to derive a quadrature formula for the direction of travel. The perhaps most common approach is the product quadrature rule. Here. a Gauss quadrature is used for 
$\mu$ and equally weighted and spaced points for $\varphi$, i.e., when using $N_q$ points, we have
\begin{equation}
\varphi_i = i \Delta\varphi \quad \text{for} \quad i=1,\ldots,N_q \quad \text{and} 
\quad \Delta\varphi = \frac{2\pi}{N_q}\;.
\end{equation}
If the Gauss quadrature for $\mu$ uses $N_q$ points, then we obtain a total of $Q = N_q^2$ possible directions. Denoting the Gauss weights as $w^G_k$ with $k = 1,\cdots,N_q$, we obtain the product quadrature weights 
\begin{align*}
    w_{k\cdot N_q +\ell} = \frac{2\pi w^G_k}{N_q}
\end{align*}
and points
\begin{align}\label{eq:spherical-coordinatesProduct}
 \mathbf \Omega_{k\cdot N_q +\ell}  = \begin{pmatrix}
           \sqrt{1-\mu_k^2}\sin(\varphi_{\ell}) \\
           \sqrt{1-\mu_k^2}\cos(\varphi_{\ell})
         \end{pmatrix} \;.
\end{align}
Other implemented quadrature methods include spherical Monte Carlo, Levelsymmetric~\cite{2004PhDT}, LEBEDEV~\cite{osti_7057084} and LDFESA~\cite{jarrel_adams}. A comparison of different quadrature sets and their approximation behaviour for $S_N$ methods can be found in \cite{camminady2021highly}.

The evolution equations for $\psi_q(t,\mathbf x):= \psi(t,\mathbf x,\mathbf \Omega_q)$ are then given by
\begin{equation}\label{eq:SNEqns}
\begin{aligned}
    \partial_{t}\psi_q(t,\mathbf x)+&\mathbf \Omega_q\cdot\nabla_x \frac{\psi_q(t,\mathbf x)}{\rho(\mathbf{x})}+\Sigma_t(t)\psi_q(t,\mathbf x) = \sum_{p=1}^{Q}w_p\Sigma_s(t,\mathbf \Omega_q\cdot\mathbf \Omega_p)\psi_p(t,\mathbf x)\;.
\end{aligned}
\end{equation}
A main disadvantage of $S_N$ methods are so called ray-effects \cite{lathrop1968ray,morel2003analysis,mathews1999propagation}, which are spurious artifacts that stem from the limited number of directions in which particles can travel. Moreover, radiation therapy applications exhibit forward-peaked scattering, 
which cannot be well-captured by classical quadrature rules. 

To allow for moderate computational costs when computing scattering terms and to efficiently treat forward-peaked scattering, we transform the nodal solution to a modal description and apply the more efficient $P_N$ methodology for scattering terms. For this, we define a truncation order $N$ and construct the matrices $\mathbf{O}\in\mathbb{R}^{Q \times (N+1)^2}$ which maps the modal onto its nodal representation and $\mathbf{M}\in\mathbb{R}^{(N+1)^2\times Q}$ which maps the nodal onto its modal representation. Such matrices can be constructed by
\begin{align*}
    \mathbf O = \left(\mathbf{m}(\Omega_k)\right)_{k=1}^{Q}\, , \text{ and } \enskip \mathbf M = \left(w_k\mathbf{m}(\Omega_k)\right)_{k=1}^{Q}.
\end{align*}
In this case, we can replace the scattering term on the right-hand side of \eqref{eq:SNEqns} by its $P_N$ counterpart. Collecting the nodal solution in the vector $\bm\psi$ then gives
\begin{equation}\label{eq:SNEqns2}
\begin{aligned}
    \partial_{t}\bm\psi(t,\mathbf x)+&\mathbf \Omega_q\cdot\nabla_x \frac{\bm\psi(t,\mathbf x)}{\rho(\mathbf{x})}+\Sigma_t(t)\bm{\psi}(t,\mathbf x) = \mathbf{O}\bm{\Sigma}\mathbf{M}\bm{\psi}\;.
\end{aligned}
\end{equation}

\section{Discretization methods}\label{sec:discretization}
\subsection{Spatial Discretization}
The KiT-RT framework is based on unstructured, cell-centered grids as spatial discretization. In the following, we restrict ourselves to a two-dimensional spatial grid, however, the notations are straight forwardly extended to three or one spatial dimensions.
A unstructured grid $\tilde {\mathbf{X}}=\menge{\mathbf X_i}_{i\in I}$ is a partition of a bounded spatial domain $\mathbf X\subset\mathbb{R}^d$. A grid cell $\mathbf X_i$ holds information about the coordinates of its centroid $\mathbf x_{i}$, its measure $A_i$, the indices of its boundary vertices, indices of its neighbor cells $N(i)$ and cell faces. The information of the cell faces is encoded in the unit-normal vector of the face dividing cell $i$ and its neighbor $j\in N(i)$, multiplied with the measure of the face and is denoted by $\mathbf n_{i,j}$. The grids used in this work are either triangular or quadrilateral, unstructured grids in two spatial dimensions. 
\subsection{Finite volume methods}
The nodal and modal methods are different approaches to discretize the velocity space of the Boltzmann equation, however all of them result in a system of transport equations, that can be solved using a finite volume scheme. Thus, we first describe a method agnostic finite volume scheme and afterwards point out the differences of the $S_N$, $P_N$ and $M_N$ based implementations. We denote the temporal variable by $t$, however the results hold for energy interpreted as pseudo-time as well.
Let $\mathbf g(t, \mathbf x)\in\mathbb{R}^m$ be the vector of conserved variables of a system of transport equations
\begin{align}\label{eq_velo_discretized_pde}
    \partial_t \mathbf g(t,\mathbf x) + \nabla_{\mathbf x}\cdot \mathbf F(\mathbf g(t,\mathbf x))= \mathbf R(t,\mathbf x,\mathbf g(t,\mathbf x)), \qquad \mathbf x\in \mathbf X,\,  t\in[0,t_f)
\end{align}
where $\mathbf F$ is the general flux function describing the solution transport, and $\mathbf R$ is a general right hand side, containing velocity discretizations of collision terms, sources and absorption terms. The main discretization strategy is to divide the spatial domain into an unstructured grid with $N_x$ cells and the time domain into $N_t$ discrete values $0=t_0<\dots<t_{N_t-1}$. We consider the solution as an average over one space-time cell
\begin{align}
   \mathbf g_i^n = \frac{1}{A_i}\int_{ \mathbf X_i}  \mathbf g \intD  \mathbf x
\end{align}
and average Eq.~\eqref{eq_velo_discretized_pde} over one space-time cell
\begin{align}
\begin{aligned}
    \frac{1}{\Delta t A_i}\int_{\mathbf X_i}\int_{t_n}^{t_{n+1}}\partial_t \mathbf g(t,\mathbf x)\intD t \intD \mathbf x& + \\
    \frac{1}{\Delta t A_i}\int_{\mathbf X_i}\int_{t_n}^{t_{n+1}}\nabla_{\mathbf x}\cdot F( \mathbf g(t,\mathbf x))\intD t \intD \mathbf x& = \frac{1}{\Delta t A_i}\int_{\mathbf X_i}\int_{t_n}^{t_{n+1}} R(t,\mathbf x, \mathbf g(t,\mathbf x))\intD t \intD \mathbf x,
\end{aligned}
\end{align}
where $\Delta t=t_{n+1}-t_n$. Solving the integrals using the Gauss theorem for the advection term and an explicit Euler scheme for the time derivative yields
\begin{align}
\begin{aligned}
    \frac{1}{\Delta t}\left(\mathbf g_i^{n+1}-\mathbf g_i^{n}\right)&+ \\ 
    \frac{1}{\Delta t A_i}\int_{t_n}^{t_{n+1}} \sum_{j\in N(i)} F(\mathbf g(t,x_{i,j}))\cdot \mathbf n_{i,j}\intD t & = \frac{1}{\Delta t}\int_{t_n}^{t_{n+1}} R(t,\mathbf x,\mathbf g(t,x_i))\intD t,
\end{aligned}
\end{align}
where $g(t,x_i)$ is the conserved variable evaluated at cell $i$ and $g(t,x_{i,j})$ is the conserved variable evaluated at the interface between cell $i$ and its neighbor $j$.
To compute the actual value of $\mathbf g_j^{n+1}$, one needs to find approximations for the flux integral. A common ansatz is of the form 
\begin{align}
    F(\mathbf g_j^n,\mathbf g_i^n)\approx  \frac{1}{\Delta t}\int_{t_n}^{t_{n+1}}F(\mathbf g(t,x_{i,j}))\cdot \mathbf n_{i,j}\intD t,
\end{align}
where the numerical flux $F(\mathbf g_j^n,\mathbf g_i^n)$ at face $(i,j)$ is approximated using the cell averaged conserved variable at cell $i$ and $j$. For transport equations, a well known numerical flux is given by the Upwind scheme~\cite{leveque}
\begin{align}\label{eq_upwind1stOrder}
     F(\mathbf g_j^n,\mathbf g_i^n)_{up} = F(\mathbf g_i^n)\cdot \mathbf n_{i,j} H(\mathbf n_{i,j}\cdot \mathbf v) + F(\mathbf g_j^n)\cdot \mathbf n_{i,j} \left( 1- H(\mathbf n_{i,j}\cdot \mathbf v) \right),
\end{align}
where $H$ is the heaviside step function and $\mathbf v$ is the transport velocity vector.
Finally, we approximate the source, absorption and collision terms using the current cell average. Thus the explicit solution iteration of a first order scheme reads
\begin{align}
\begin{aligned}
   \mathbf g_i^{n+1}=
    \mathbf g_i^{n}-\frac{\Delta t}{ A_i} \sum_{j\in N(i)} F(\mathbf g_j^n,\mathbf g_i^n)_{up} + \Delta t R(t,\mathbf x_i,\mathbf g_i^n).
\end{aligned}
\end{align}
Due to the fact that the scattering term $R$ is commonly stiff, implicit-explicit (IMEX) schemes can be used to remove influences of scattering from time step restrictions. If we assume a linear scattering term, that is with a given matrix $\mathbf{R}_i^{n+1}$ we have $R(t^{n+1},\mathbf{x}_i,\mathbf g_i^n) = \mathbf{R}_i^{n+1}\mathbf g_i^n$, the IMEX scheme reads
\begin{align}
\begin{aligned}
   (\mathbf I - \Delta t \mathbf{R}_i^{n+1})\mathbf g_i^{n+1}=
    \mathbf g_i^{n}-\frac{\Delta t}{ A_i} \sum_{j\in N(i)} F(\mathbf g_j^n,\mathbf g_i^n)_{up}.
\end{aligned}
\end{align}

\subsection{Second order finite volume schemes}
The KiT-RT solver provides the option to evaluate the space and time discretizations using second order accurate schemes. To this end, we use a Heun scheme for the temporal discretization and a second order upwind flux for the numerical flux~\cite{barthJespersen}. Whereas, first order spatial fluxes assume a constant solution value $\mathbf g_i^n$ in a cell $i$,
a second order upwind scheme is based on a linear reconstruction of the conserved variable. Therefore the inputs  $\mathbf g_i^n$ and  $\mathbf g_j^n$ to the numerical flux of Eq.~\eqref{eq_upwind1stOrder} are replaced by
\begin{align}
    \tilde{\mathbf g}_i^n &= \mathbf g_i^n + \Psi_i \left(\nabla_{\mathbf{x}}\mathbf g_i^n\cdot \mathbf{r}_{i,j} \right),\\
    \tilde{\mathbf g}_j^n &= \mathbf g_j^n + \Psi_j \left(\nabla_{\mathbf{x}}\mathbf g_j^n\cdot \mathbf{r}_{j,i} \right),
\end{align}
where $\mathbf{r}_{i,j}$ is the vector pointing from cell centroid $\mathbf x_i$ of cell $i$ to the interface midpoint between cells $i$ and $j$, and $\Psi_i$ is the flux limiter for cell $i$. This reconstruction is formally second order accurate on regular grids~\cite{AftosmisGaitonde} assuming exact evaluation of the gradient $\nabla_{\mathbf{x}}\mathbf g_i^n$. The gradient of the conserved variable $\mathbf g_i^n$ is evaluated using the Green-Gauss theorem with interpolated solution values at the  cell interfaces,
\begin{align}
    \nabla_{\mathbf{x}}\mathbf g_i^n\approx \frac{1}{A_i}\sum_{j\in N(i)}\frac{1}{2}\left(\mathbf g_i^n + \mathbf g_j^n\right)\cdot\mathbf{n_{i,j}}.
\end{align}
Second or higher order upwind spatial discretizations require the use of flux limiters in order to prevent the generation of oscillations in shock regions and to achieve a monotonicity preserving scheme. In the KiT-RT package, the Barth and Jespersen~\cite{barthJespersen} limiter as well as the Venkatakrishnan limiter~\cite{venkatakrishnan} are implemented. Exemplary, we show the computation of the  Barth and Jespersen limiter at cell $i$
\begin{align}
    \Psi_i = \min_j
\begin{cases}
\min(1,\frac{\mathbf{g}_{\max}-\mathbf{g}_i}{\Delta_2}), &\text{ if } \Delta_2>0\\
\min(1,\frac{\mathbf{g}_{\min}-\mathbf{g}_i}{\Delta_2}), &\text{ if } \Delta_2<0\\
1, &\text{ else}
\end{cases},
\end{align}
where we have
\begin{align}
    \Delta_2 &= \frac{1}{2}\nabla_{\mathbf x}\mathbf g_i^n\cdot\mathbf r_{i,j},\\
    \mathbf{g}_{\max} &= \max(\mathbf{g}_{i},\mathbf{g}_{j}),\\
    \mathbf{g}_{\min} &= \min(\mathbf{g}_{i},\mathbf{g}_{j}).
\end{align}
The second order Heun scheme for temporal discretization is a two step Runge-Kutta scheme with the iteration formula
\begin{align}
\begin{aligned}
   \mathbf g_i^{*}&=
    \mathbf g_i^{n}-\frac{\Delta t}{ A_i} \sum_{j\in N(i)} F(\mathbf g_j^n,\mathbf g_i^n)_{up} + \frac{\Delta t}{ A_i} R(\mathbf g_i^n),\\
    \mathbf g_i^{**}&=
    \mathbf g_i^{*}-\frac{\Delta t}{ A_i} \sum_{j\in N(i)} F(\mathbf g_j^*,\mathbf g_i^*)_{up} + \frac{\Delta t}{ A_i} R(\mathbf g_i^*), \\
     \mathbf g_i^{n+1}& = \frac{1}{2}\left(\mathbf g_i^{n} +  \mathbf g_i^{**}\right),
\end{aligned}
\end{align}
which is based on the implicit trapezoidal integration method.
\subsection{Numerical Fluxes}
In the following we adapt the introduced numerical methods to the method specific notation and present the detailed implementation. 
The space and time averaged conservative variables $\mathbf g_i^n$ for the nodal discretization are given by the vector of the radiation flux $\bm{\psi} =[\psi_1,\dots,\psi_{N_q}]^T$ evaluated at the quadrature points and for the modal discretization by the moment vector $\mathbf{u}$.
The different methods are distinguishable by their numerical flux function. 
The corresponding numerical flux for the $S_N$ method is given by
\begin{align}
  F(\mathbf{ {\psi}_i^n}) =\mathbf{\Omega}\otimes\frac{\mathbf {\psi}_{i}^n}{\rho(\mathbf{x_i})}
\end{align}
and the corresponding upwind flux reads
\begin{align}\label{eq_upwind1stOrder2}
     F(\psi_j^n,\psi_i^n)_{up} =  \mathbf{\Omega}\cdot \mathbf n_{i,j}\frac{\mathbf {\psi}_{i}^n}{\rho(\mathbf{x_i})} H(\mathbf n_{i,j}\cdot \mathbf{\Omega}) + \mathbf{\Omega}\cdot \mathbf n_{i,j}\frac{\mathbf {\psi}_{j}^n}{\rho(\mathbf{x_j})} \left( 1- H(\mathbf n_{i,j}\cdot \mathbf{\Omega}) \right).
\end{align}
The numerical flux for the $P_N$ method reads
\begin{align}
  F(\mathbf{ \mathbf{u}_i^n}) = \left[\mathbf A_1 \mathbf {u}_{i}^n,\mathbf A_2 \mathbf {u}_{i}^n,\mathbf A_3 \mathbf {u}_{i}^n\right]^T,
\end{align}
where $A_i$ are the flux Jacobians emerging from the spherical harmonics recursion scheme.To evaluate the numerical flux with an upwind scheme, we decompose the flux Jacobians in their positive and negative definite parts,
\begin{align}
    A_l = A_l^+ + A_l^-,\qquad l=1,\dots,d
\end{align}
Then the numerical flux is given by
\begin{align}
\begin{aligned}
     F(\mathbf{u}_i^n,\mathbf u_j^n)_{up} = \sum_{l=1}^d &\left(A_l^+\frac{\mathbf{u}_i^n}{\rho(\mathbf x_i)} + A_l^- \frac{\mathbf{u}_j^n}{\rho(\mathbf x_j)}\right)n_l  H(n_l) +\\ &\left(A_l^-\frac{\mathbf{u}_i^n}{\rho(\mathbf x_i)} + A_l^+\frac{\mathbf{u}_j^n}{\rho(\mathbf x_j)}\right)n_l\left( 1- H(n_l)\right).
\end{aligned}
\end{align}
In contrast to the $P_N$ method, the flux function of the $M_N$ method cannot be expressed as a matrix multiplication, but reads
\begin{align}
   F(\mathbf{ \mathbf{u}_i^n})=  \int_\mathbb{S}^2 \mathbf{\Omega}\otimes \mathbf m(\mathbf{\Omega})\psi_{\mathbf u^i_n}(\mathbf{\Omega})\intD \mathbf{\Omega},
\end{align}
where $\psi_{\mathbf u^i_n}$ is the reconstructed radiation flux density of the minimal entropy closure at the cell averaged moment~${\mathbf u^i_n}$.
Using a quadrature rule for the velocity integral discretization and a numerical flux for every quadrature point, we arrive at the kinetic numerical flux
\begin{align}
    \begin{aligned}
     F(u_j^n,u_i^n)_{up} = \\  
     \sum_{q=1}^Q w_q \mathbf m_q \mathbf{\Omega}_{q}\cdot \mathbf n_{i,j} &\left[\frac{\psi_{\mathbf u_i^n,q}}{\rho(\mathbf{x}_i)}    H(  \mathbf{\Omega}_{q}\cdot \mathbf n_{i,j}) + \frac{\psi_{\mathbf u_j^n,q}}{\rho(\mathbf{x}_j)}  \left( 1- H( \mathbf{\Omega}_{q}\cdot \mathbf n_{i,j}) \right) \right].
\end{aligned}
\end{align}
Note, that the updated solution of the $M_N$ method must still be a feasible moment for the minimal entropy closure of Eq~\eqref{eq_entropyOCP}. To ensure this, one must either employ a flux-limiter~\cite{KRISTOPHERGARRETT2015573}, construct a realizability reconstruction~\cite{kusch2018maximumprinciplesatisfying} or employ the regularized entropy closure formulation.\\
The numerical framework supports the usual Neumann and Dirichlet boundary conditions.
\section{Software architecture}
\label{sec:softwareArch}
\begin{figure}
    \centering
    \includegraphics[width=1\linewidth]{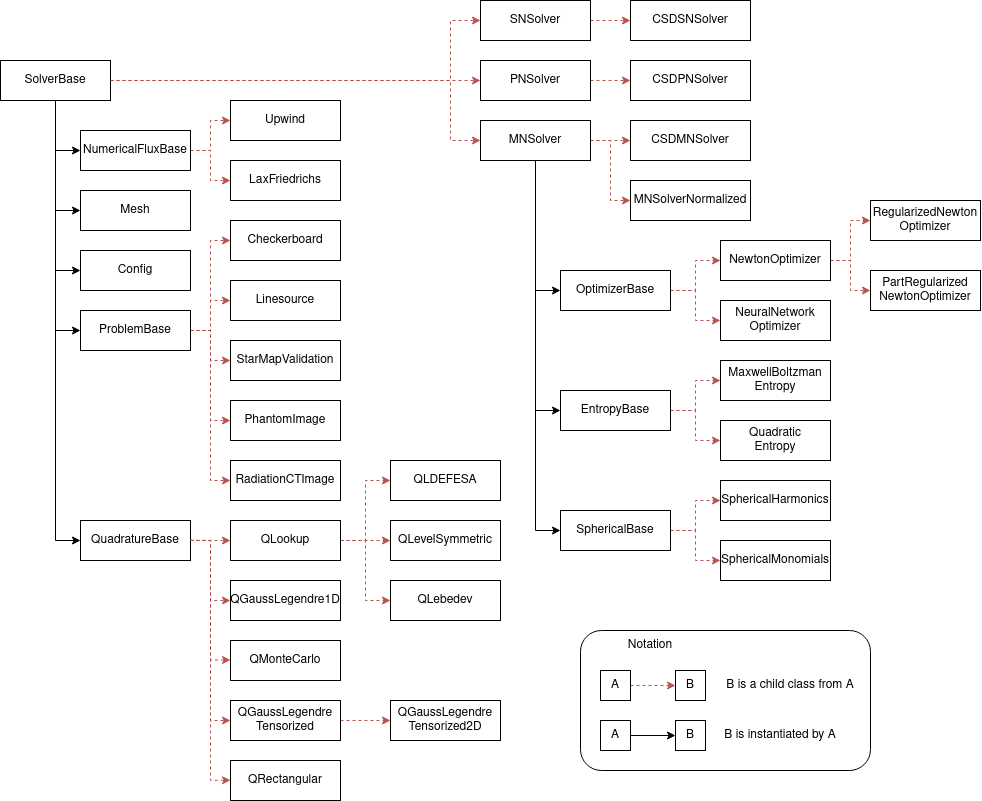}
    \caption{Class and inheritance structure of the virtual SolverBase Class. Each instantiated Solver has  class members and routines specific to its numerical structure.}
    \label{fig_class_diagram}
\end{figure}

The design principle of the KiT-RT software package is focused on efficient implementation, high re-usability of its components and ease of extension. It contains a set of efficient numerical solvers for radiation transport, which are constructed of basic, reusable building blocks. These building blocks can be freely arranged to implement new solvers or tools for completely different applications.
On the other hand, KiT-RT is equipped with an easy to use command line interface based on readable configuration files, which allow easy manipulation of the solvers.
Thus the software is attractive for developers, who want to experiment with the framework and build their own numerical solvers as well as users and application engineers, who want to gain experimental insights without directly interfering with the codebase.
\\
KiT-RT is implemented in modern C++ and uses mainly polymorphism for its construction. In the following, we present the class structures used to build the numerical solvers and explain the used building blocks, which are displayed in Fig.~\ref{fig_class_diagram}.
Most building blocks consist of a virtual base class, which contains a static factory method to build an instance of the concrete derived class, defined by the given configuration details. Furthermore, the virtual base class defines the interface of this building block with other parts of the KiT-Framework. 

\subsection{Solver Class}
\label{sec:solverClass}
The virtual {SolverBase} class is the basic blueprint for all time or energy dependent  finite volume solvers of the KiT-RT framework. It holds an instance of the {Config}, NumericalFlux, ProblemBase, QuadratureBase and Mesh class. \\
It controls the screen, log and volume output of the solver. The screen output provides instantaneous feedback of the solver state via the command line and gives information on the current iteration, the total mass of the system, the residual of the radiation flux as well as the flow field and whether logs and volume outputs have been written to file. The file log carries the same information as the screen output in a tabular format. Lastly, the volume output consists of vtk files with solver and problem specific solution data. The output data can be specified in the solver configuration. \\
\begin{algorithm}
 \DontPrintSemicolon
 \caption{Execution of the solver}
\SetAlgoLined
\SetKwInOut{Input}{Input}
Prepare outputs for Screen, Log and Volume\;
Solver specific pre-processing\;
\For {$i = 1,\dots,n_{tf}$}{
    \For {$l = 1,\dots,k_{rk}$}{
    Runge-Kutta pseudo iteration pre-processing\;
    Update numerical fluxes\;
    Perform finite volume step\;
    Update Runge-Kutta intermediate solution\;
    }
    Iteration post-processing\;
    Compute Runge-Kutta solution\;
    Write Screen, Log and Volume output as configured\;
 }
Post-processing of output\;
\label{alg_basicSolve}
\end{algorithm}
The method  {Solve()} of the SolverBase class drives the execution of all derived solvers by iterating over the time discretization of the numerical methods described in Section~\ref{sec:discretization}. This main time-iteration is displayed in Algorithm~\ref{alg_basicSolve}.
Each command is specified in the derived solver classes such as the $P_N$ solver and does not induce any additional communication overhead for the parallelization architecture.
The class PNSolver inherits from SolverBase and does not own additional instances of other custom building blocks and overwrites the sub-routines of Algorithm~\ref{alg_basicSolve} for the $P_N$ equation specific numerical method, which allows for runtime solver assembly. It's child class is the CSDPNSolver, which is the implementation of the $P_N$ based continuous slowing down solver, that overwrites the solver-preprocessing routines for the continuous slowing down specific energy transformation. The $P_N$ based solvers produce radiation flux and moments as output.

The class SNSolver adapts the sub-routines of Algorithm~\ref{alg_basicSolve} for the ordinate based numerical methods and is the parent class of the CSDSNSolver. $S_N$ based solvers produce the radiation flux as output. 

Lastly,  the MNSolver class contains the implementation of the $M_N$ numerical method and holds the module SphericalBase, which controls the choice of basis functions $m(v)$ of the velocity space, the module EntropyBase, that controls the choice of entropy functional for the entropy closure and lastly the module OptimizerBase, which controls the choice of numerical optimizer used to compute the entropy closure. The class CSDMNSolver inherits from the MNSolver class and analogously overwrites the sub-routines of  Algorithm~\ref{alg_basicSolve} for the continuous-slowing down equations. $M_N$ based solvers produce the radiation flux, moments and dual variable of the entropy closure as output. 
\subsection{Mesh Class}
The mesh class handles the computational meshes of the spatial discretization of the underlying differential equation. It can handle $1D$ meshes and $2D$ unstructured triangular and quadrilateral meshes in the {SU2}~\cite{SU2} mesh format. The mesh class keeps a record of all geometry and adjacency information required for the finite volume methods with first and second order fluxes.
\subsection{Computational problem class}\label{sec:problemClass}
The problem class handles the setup of computational problems and test cases. It sets the initial conditions for the solution of the numerical solver and manages space, time or energy dependent material properties for the solver. The abstract class problem base holds pointers to the Mesh and Config classes and creates instances of specific problems depending on the chosen configurations. Each implemented problem has two child classes, one for the ordinate based and one for moment based solvers. The moment based problem classes compute the moments of the initial condition and sources of the corresponding kinetic densities specified in the ordinate based problem class.
\\
The solver framework comes with a number of pre-implemented test cases and functionalities. This includes standard 1D and 2D test cases such as line source and checkerboard for the radiative transfer solvers, as well as isotropic and directed sources with different background media which can be loaded from a user-supplied image file, for the continuous slowing down solvers. Custom test cases can be easily added by the user, based on the provided examples and our modular approach.

\subsection{Quadrature Class}
The virtual quadrature base class creates instances of specific numerical quadratures using its static factory method. 
The quadratures are intended to integrate over the velocity space of the Boltzmann equation, however, they can be applied to other use-cases as well. 
The implemented quadratures are distinguished by the dimension of the integrated velocity space and the integration area. Each quadrature has a specifiable order and manages the integration points in Cartesian and spherical coordinates as well as the corresponding quadrature weights. By default, the quadrature rules integrate over the unit sphere, but the integration region can be scaled.
\subsection{IO/Use of Config Files}
The KiT-RT solver is a command line interface based program and takes one argument, the configuration file. This file is parsed and the specified modules of the KiT-RT framework are arranged for a solver instance or another custom tool.
The configuration file is a document containing option specifications of the form CONFIG\_OPTION=VALUE. A solver configuration contains information about file input and output, where the location of the mesh file, the volume output files and the log files are specified. Then, the computational problem and the problem specific parameters, e.g. scattering coefficient, final time, spatial dimension and boundary conditions, are set. Next, the solver specific options are set. In the example of an $M_N$ solver, the choice of velocity basis, the maximal degree of the basis functions, the CFL number, spatial integration order, entropy functional, optimizer, quadrature and quadrature order are set.
Finally, quantities for screen, volume and log output are specified along with their output-frequency.
Example configuration files for the numerical results of Section~\ref{sec:validation} can be found in the Github Repository~\url{https://github.com/CSMMLab/KiT-RT}.
\subsection{Practices of modern software development}

The entire solver and associated documentation is put under the version control system  git \cite{chacon2014pro} to greatly enhance collaborative workflows. Additionally, the web-hosting service GitHub is used to provide global access to the code which is licensed under the open-source MIT license.\\
To further improve collaborations, the service also acts as a central host for progress and issue tracking, deployment and maintaining code integrity. The latter is obtained through automated testing in terms of unit test, which ensure the validity of smaller code instances such as functions or classes by testing predefined inputs against their expected result, or regression tests which validate the correctness of the solver as a whole based on small test problems and compare obtained results to reference solutions.
These test are automatically executed every time code changes are submitted to the main development branches or if a merge request is opened.\\
If any of the automated tests fail for a new submit it is rejected for merging, ensuring code integrity and quality on the important development streams at all times.  
Combining the test information, we can further define metrics such as a test coverage describing the percentage of code lines validated by any form of testing and ultimately helps building trust in the code framework. For the KiT-RT framework the test coverage is reported to the coveralls.io service at \url{https://coveralls.io/github/CSMMLab/KiT-RT}.
The KiT-RT framework features relatively modest software dependencies, but nevertheless being able to build and run the code correctly can be troublesome on many systems. To circumvent this issue, we provide a pre-configured build environment through the containerization engine Docker \cite{merkel2014docker}. These so-called Docker containers have been developed for consistent software development and deployment and work as isolated instances with a minimal software stack comparable to lightweight virtual machines. The respective specialized docker image is also publicly available at \url{https://hub.docker.com/r/kitrt/test}.\\
As also mentioned previously, GitHub can also be used for the deployment of precompiled software packages and the associated documentation. The documentation is automatically is automatically generated as part of a complete software build. It is written in the reStructuredText Markup language and uses the documentation framework Sphinx \cite{brandl2021sphinx}, which compiles the Markup files to a series of linked HTML files or in other words a local website. To make the website itself publicly available it is hosted by ReadTheDocs \footnote{\url{https://readthedocs.org/}} service under the URL \url{https://kit-rt.readthedocs.io}.\\
With all these tools in mind, the development workflow can be described as follows:\\
Starting on GitHub, each developer can create a new branch based on the development branch or fork the entire KiT-RT framework to obtain a personal workspace. After the developers have added their changes, they can file a merge request that will automatically be tested by the continuous integration processes and a core developer will perform a code review of all changes. Provided all test succeed and the core developer is satisfied with the added/changed code quality, it will be merged into the development branch. If enough new features have been added into the development branch, it will be merged to the master branch and the software will obtain a new version number (major or minor).
\section{Parallel Scaling}
In the following we investigate the parallel performance of the three base solver implementations $S_N$, $P_N$ and $M_N$, where we follow~\cite{parallel_metrics} for the brief review of parallel scalings.
The speedup of a parallel algorithm is defined as 
\begin{align}\label{eq_parallel_speedup}
    S(n,p)=\frac{T^*(n)}{T(n,p)},
\end{align}
where $T^*(n)$ is the execution time of the best inherently serial algorithm with input size $n$ and $T(n,p)$ the time for the parallel implementation with $p$ processing workers and input size $n$. In general the best serial algorithm may be different than the parallel algorithm, however in our application case, the finite-volume discretization scheme does not change for serial implementation. \\
In theory the best possible speedup is linear~\cite{FABER1986259}, i.e., $S(n,p)=p$, thus the measure of parallel efficiency is 
\begin{align}
E(n,p)=\frac{S(n,p)}{p}.
\end{align}
In practise the speedup and parallel efficiency of an algorithm is limited by spawning and communication overhead of the parallel workers as well as the fraction of inherently serial code $f$, that exists in any algorithm. Thus the upper bound for the speedup is given by Amdahl~\cite{Gustafson2011}
\begin{align}
     S(n,p)\leq\frac{1}{f + (1-f)/p}
\end{align}
For larger input sizes, the fraction of inherently serial code $f$ typically decreases, which enables the use of highly parallel implementations. Two common approaches to measure the parallel performance are given by the strong and weak scaling approach, where the former describes an experiment, where for fixed input size $n$ the amount of parallel workers $p$ is increased and the their timing is measured, which directly results in the speedup of Eq.~\eqref{eq_parallel_speedup}. The latter increases the input size $n$ proportionally to the worker count $p$. A perfectly parallel algorithm would have a constant parallel time $T(n,p)$.

The philosophy of the parallel implementation of the solvers of this work are based on the independence of the performed computations of the grid cells. As Algorithm~\ref{alg_basicSolve} displays, during one (pseudo) time iteration of a solver, a set of instructions are calculated. Each instruction can be carried out independently for each grid cell and only in between two instruction sets, communication between parallel workers need to be established. Therefor, the parallel implementation spawns a set of parallel workers with shared memory access and distributes the spatial grid among them to carry out the current instruction. The input size $n$ is thus given by the number of grid cells of the spatial discretization.
\begin{figure}[htp!]
    \centering
    \includegraphics[width=0.48\linewidth]{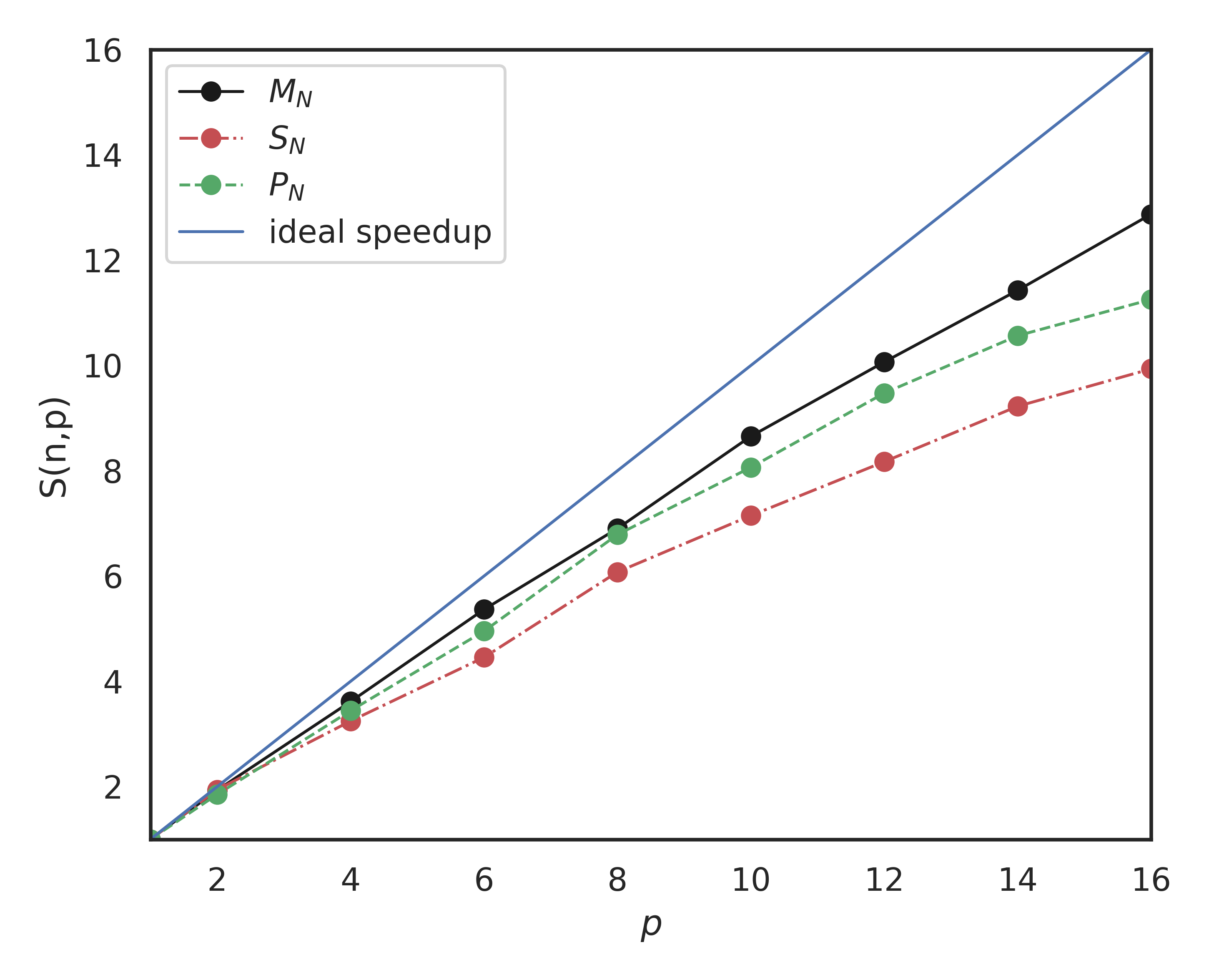}
        \includegraphics[width=0.48\linewidth]{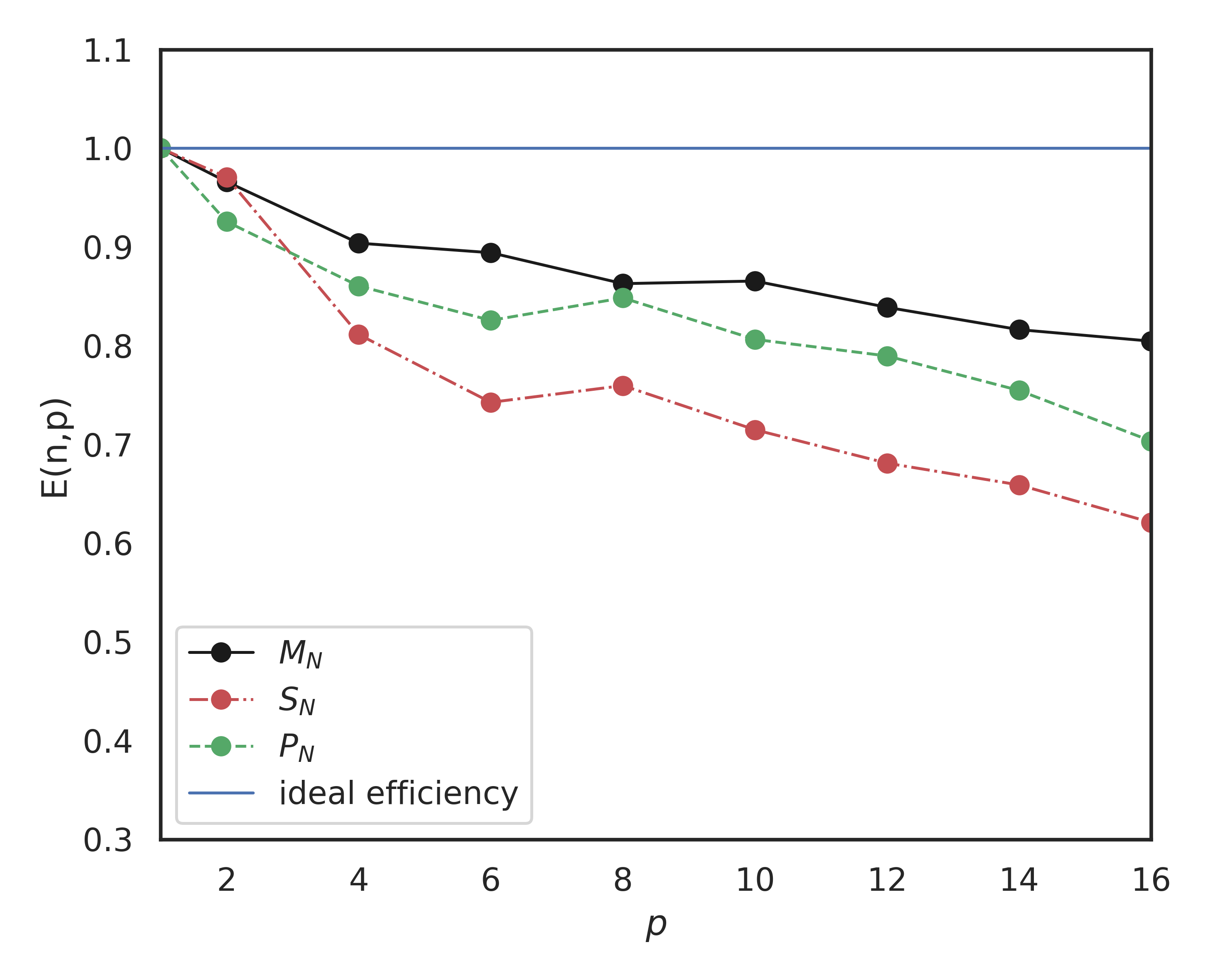}
    \caption{Strong parallel scaling (left) and parallel efficiency (right) for the $M_N$, $S_N$ and $P_N$ solver}
    \label{fig_strong_scaling}
\end{figure}
We perform a strong parallel scaling study for the implementation of the $M_N$, $S_N$ and $P_N$ solver on the Linesource test case, as described in Section~\ref{sec:inhom_linesource}, with a fixed unstructured triangular mesh of size $n=578290$ and varying number of shared memory parallel workers $p$. We choose $p=1,\dots,16$, furthermore, the solvers allocated memory does not exceed the systems memory.
Figure.~\ref{fig_strong_scaling} shows a comparison of the solvers parallel scalings and efficiency. It is apparent, that the $M_N$ solver enjoys the highest speedup even for a high parallel worker count, while the $P_N$ and $S_N$ solver experience diminishing returns for more than $p=12$ workers. 
The performance of the continuous slowing down solver implementations follows the corresponding base solver performance, since the same  spatial, velocity and (pseudo) temporal discretizations are used. 

\section{Validation}\label{sec:validation}

For validation and a comparison of the implemented solvers, we consider a selection of the test cases provided within the problem class (\ref{sec:problemClass}). 
The $S_N$ solver is validated with a comparison with Kinetic.jl~\cite{Xiao2021_kinetic,xiao2021flux}.
The continuous slowing down solvers are further compared to a reference Monte Carlo solution computed using TOPAS \cite{perl2012topas} as well as the validated spherical harmonics solver StaRMAP \cite{seibold2014starmap}.
\subsection{Inhomogeneous linesource}
\label{sec:inhom_linesource}
In the following, we compare the numerical results of our framework to a Monte Carlo solution computed using TOPAS \cite{perl2012topas} as well as the staggered-grid spherical harmonics solver StaRMAP \cite{seibold2014starmap}. The problem considered is an inhomogeneous linesource testcase, which extends the classical linesource benchmark \cite{ganapol1999homogeneous,ganapol2008analytical} to a steady-state but energy dependent setting. As background density, a piece-wise constant function $\rho(\mathbf x) = 1 + 4\cdot\mathbbm{1}_{X_u}(\mathbf x)$ for $\mathbf{x}\in [0,1]^2$ is chosen. Here $\mathbbm{1}_{A}:X\rightarrow \mathbb{R}$ denotes the indicator function for the set $A$. The upper part of the spatial domain on which we prescribe a reduced density is defined as $X_u = [0,1]\times [0.56,1]$. At a maximal energy of $E_{\mathrm{max}}=1$, a particle beam is positioned in the center of the spatial domain $\mathbf{x}_0 = \frac12(1,1)^T$, which is modelled as
\begin{align*}
        \psi(E_{\mathrm{max}},\mathbf{x},\mathbf{\Omega}) =& \frac{1}{2\pi\sigma^2}\exp\left(-\frac{\Vert \mathbf{x} - \mathbf{x}_0 \Vert^2}{2\sigma^2} \right)\\
        \psi(E_{\mathrm{max}},\mathbf{x},\mathbf{\Omega}) =& 0 \quad \text{ for } \mathbf{x}\in\partial X.
\end{align*}
Here a standard deviation of $\sigma = 0.01$ is chosen to obtain a sharp particle beam in the center. The spatial grid for all deterministic methods is a structured rectangular grid with $300^2$ cells. Due to the functionality of the Monte-Carlo software, we use a three-dimensional grid and project the $x_3$-domain onto the $x_1-x_2$ plane. To allow for feasible costs, the Monte-Carlo method uses a coarser grid resolution of 100 spatial cells per dimension and 100000 Monte-Carlo runs are computed to reduce statistical noise. The $S_N$ solver uses a product quadrature rule of order $20$ for the streaming step and spherical moments up to order $8$ to compute scattering terms. Similarly, the $P_N$ solver employs spherical moments up to order $8$. The time step restriction of all deterministic methods picks a CFL number of $0.7$. All methods are second order in space and first order in time.

\begin{figure}[htp!]
    \centering
    \includegraphics[width=0.48\linewidth]{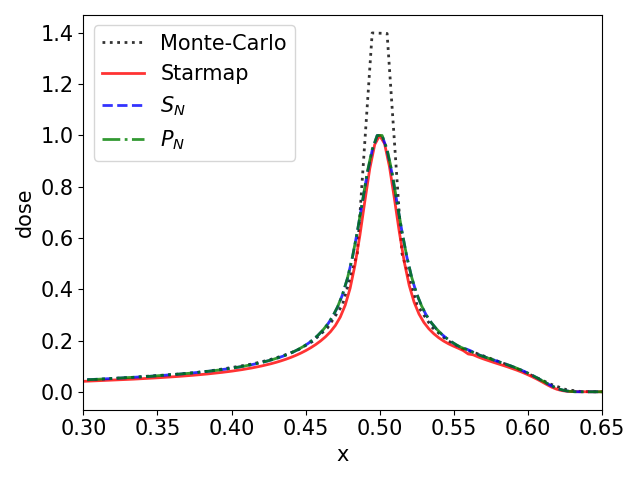}
    \includegraphics[width=0.48\linewidth]{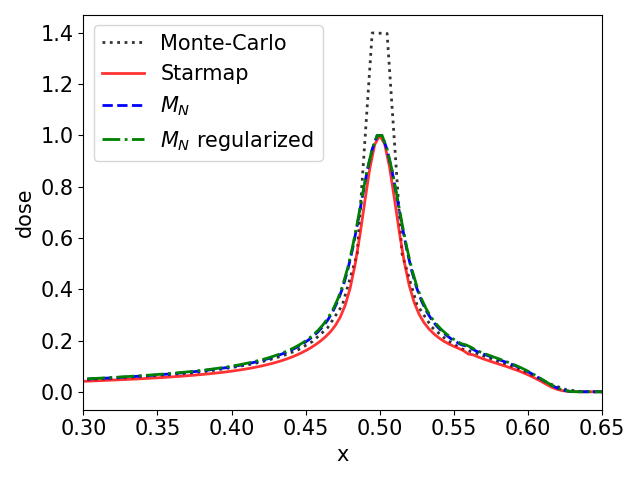}
    \caption{Comparison of simulation results of deterministic and stochastic methods.}
    \label{fig:DoseCutLS}
\end{figure}

The resulting dose profile is plotted in Figure~\ref{fig:DoseCutLS} along the $x_2$-axis in the interval $x_2\in[0.3,0.65]$. It is observed that the dose drops in the transition between densities $1$ and $5$. All methods show a similar behaviour and (except for the center region) agree well with the Monte-Carlo results. Note that the increased Monte-Carlo dose results from the coarser mesh, leading to a coarser initial particle distribution. Moreover, it is observed that the regularized $M_N$ method seems to coincide with its non-regularized counterpart.

\subsection{Checkerboard}
\begin{figure}[htp!]
    \centering
    \includegraphics[width=0.48\linewidth]{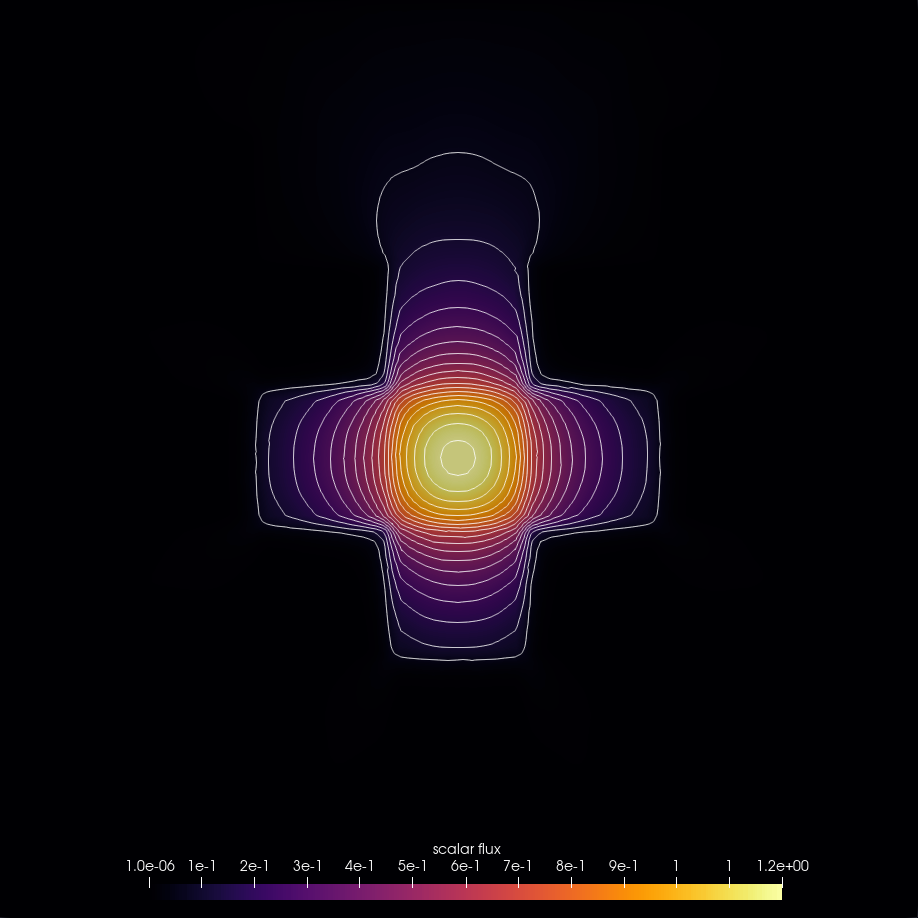}
    \includegraphics[width=0.48\linewidth]{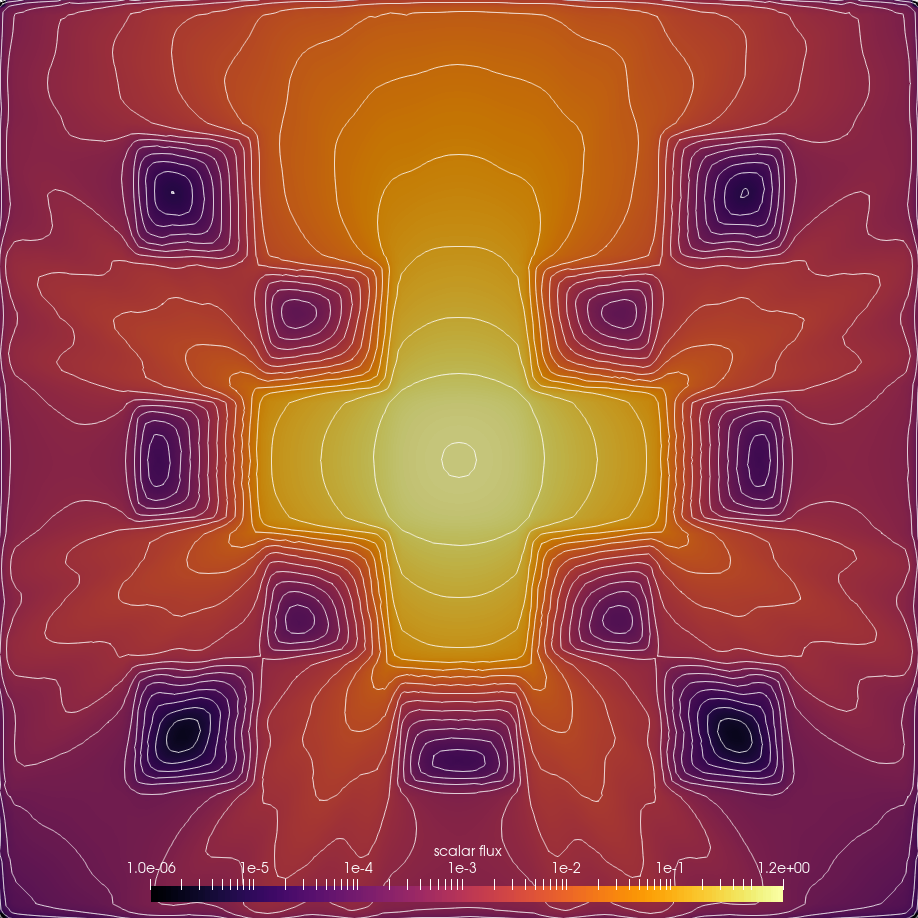}
    \caption{Simulation results for the S$_{10}$ solver in linear scale and log scale.}
    \label{fig_res_checker2d_SN}
\end{figure}

\begin{figure}[htp!]
    \centering
    \includegraphics[width=0.48\linewidth]{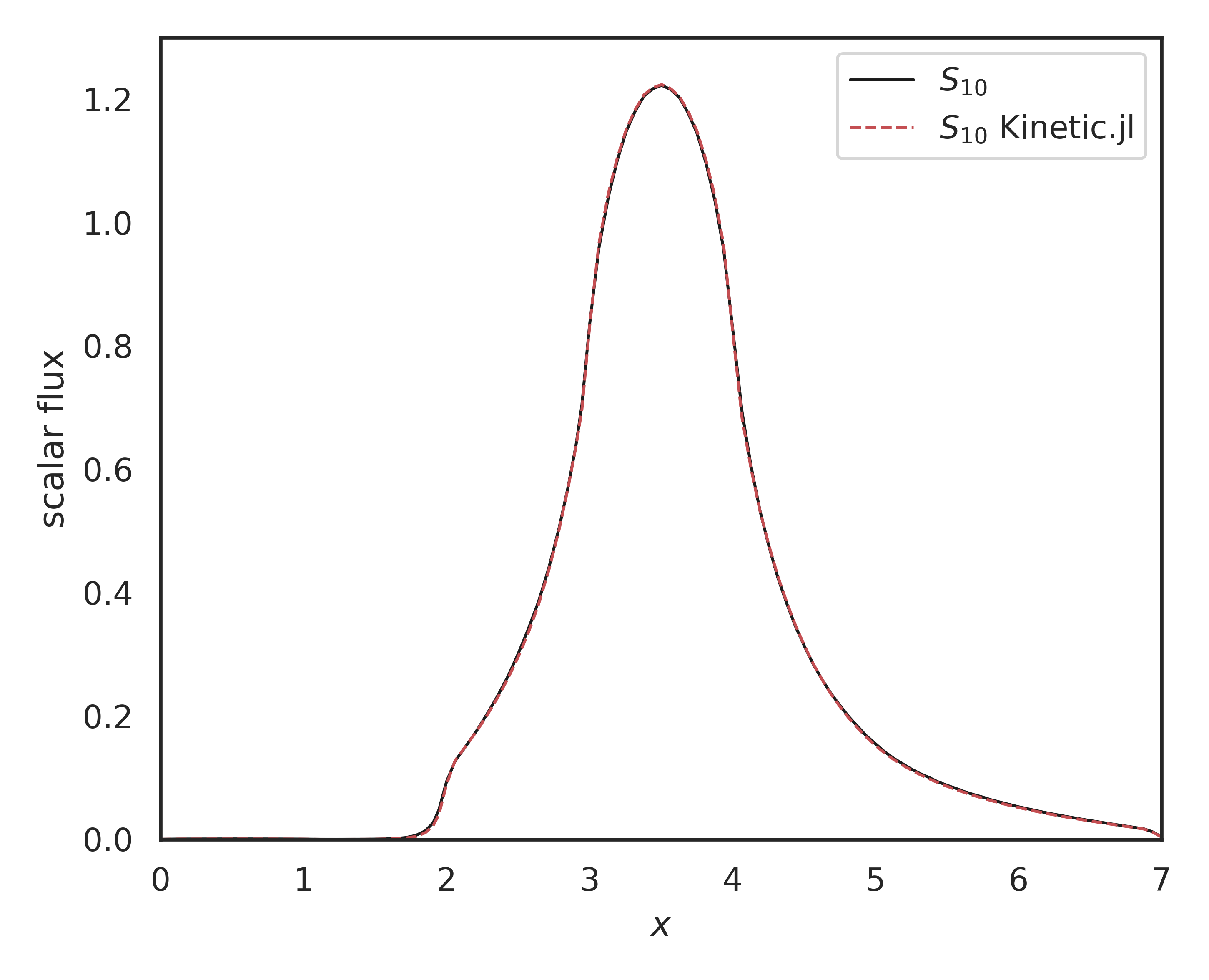}
    \includegraphics[width=0.48\linewidth]{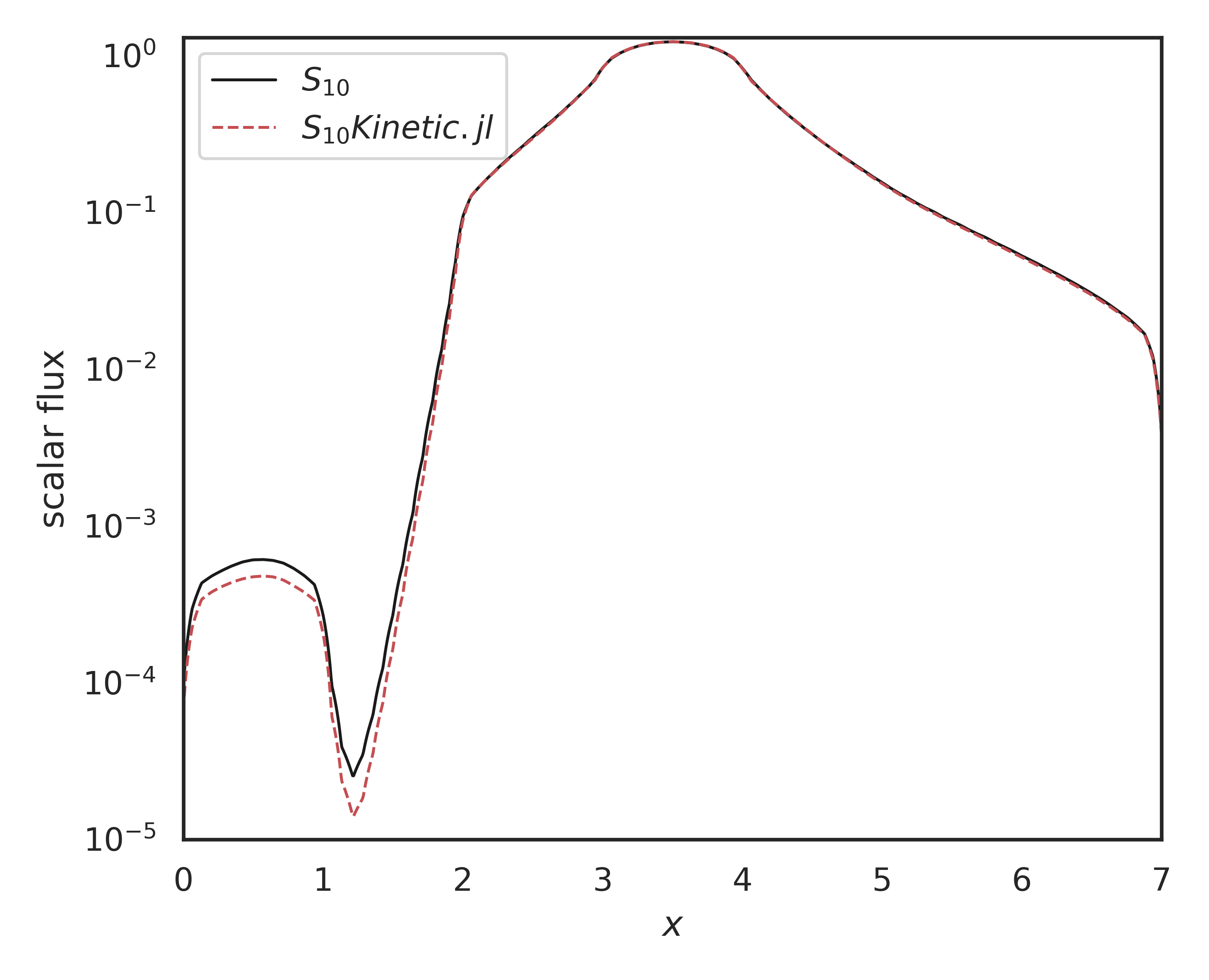}
    \caption{Vertical cross section through the solution of the S$_{10}$ method. Comparison of the KiT-RT and kinetic.jl packages.}
    \label{fig_xs_checker_SN}
\end{figure}

\begin{figure}[htp!]
    \centering
    \includegraphics[width=0.48\linewidth]{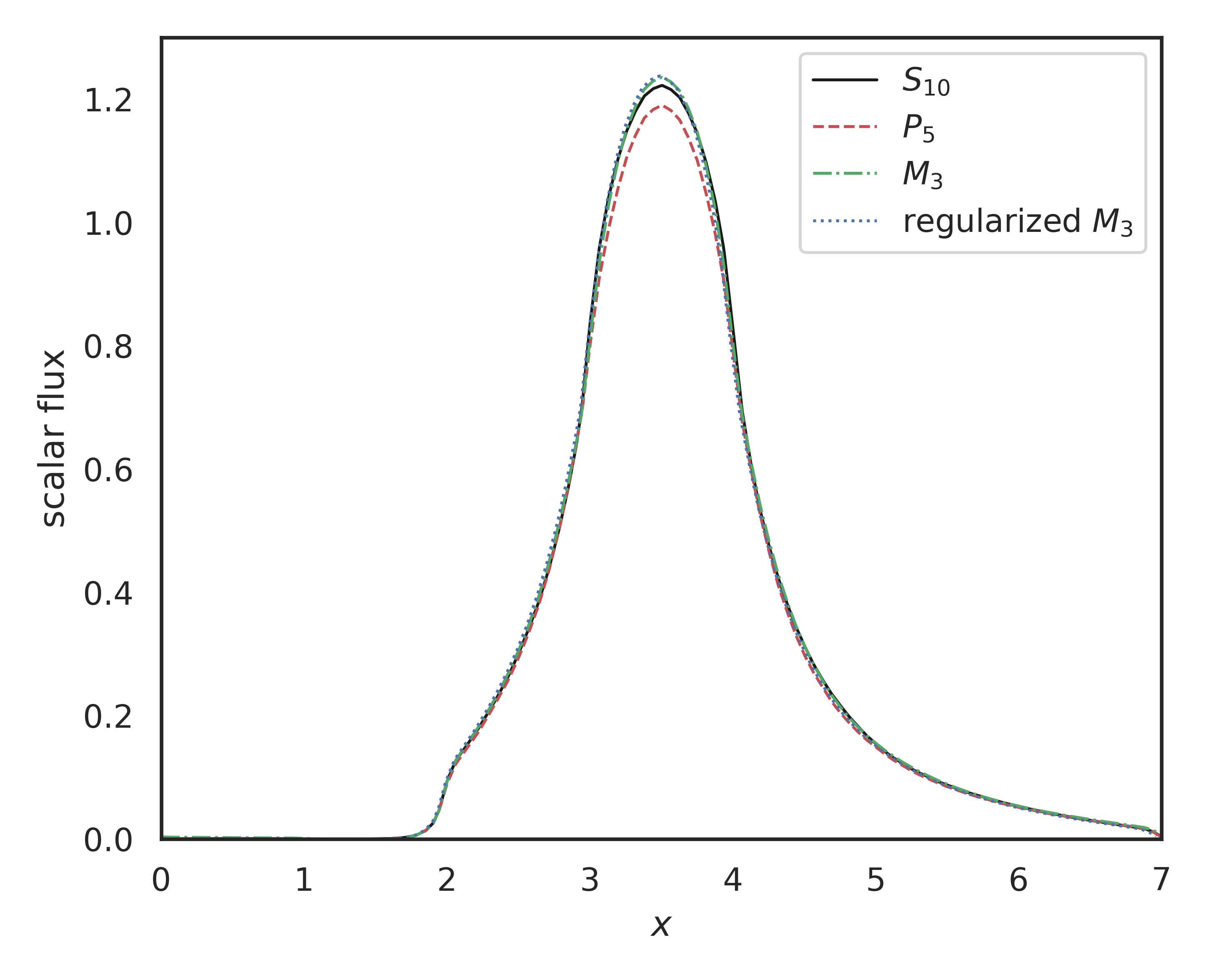}
    \includegraphics[width=0.48\linewidth]{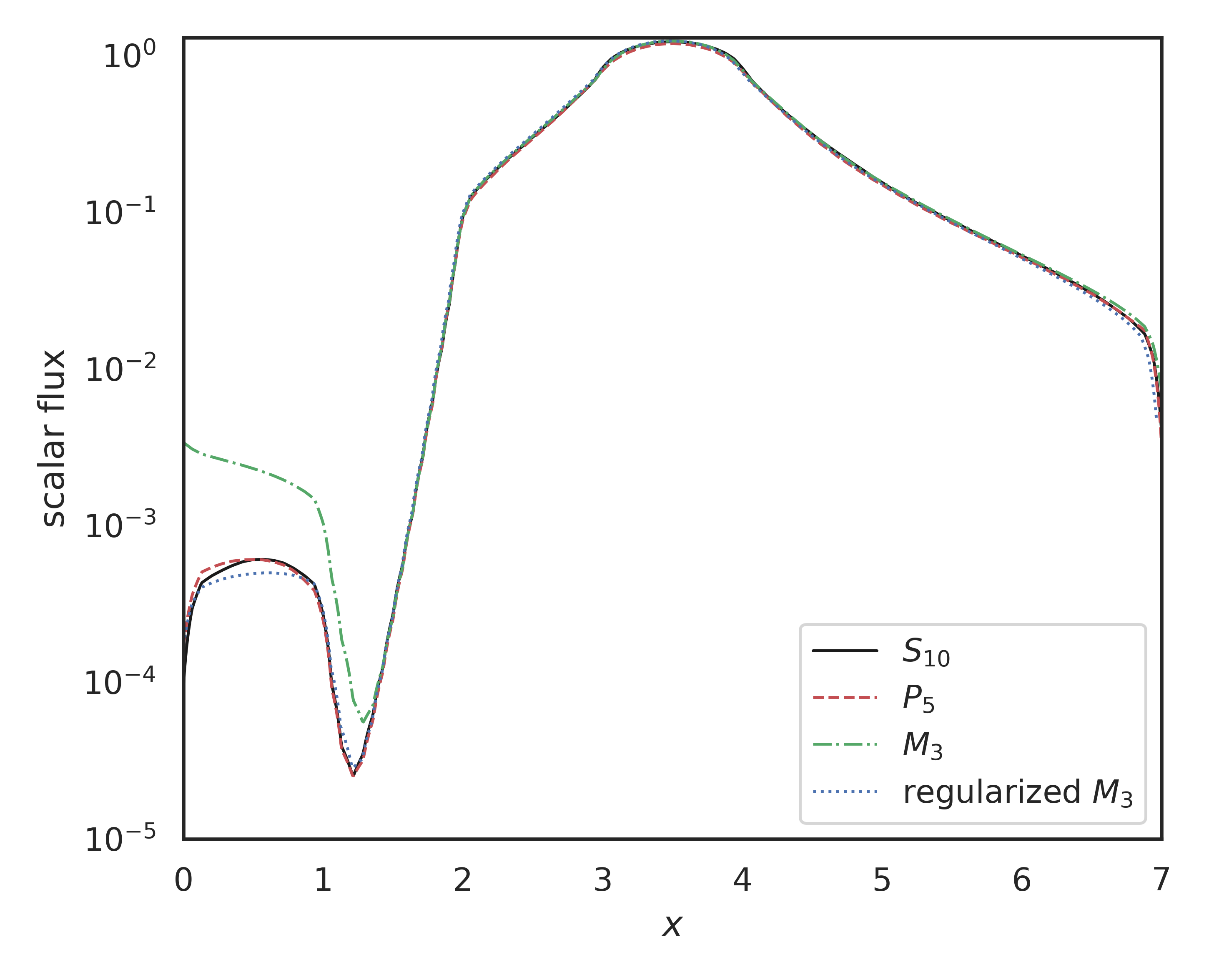}
    \caption{Vertical cross section through the solution of the checkerboard test case. Comparison of the  $S_{10}$, $P_5$ ,$M_3$ and regularized $M_3$ solver .}
    \label{fig_xs_checker_solvers}
\end{figure}

\begin{figure}[htp!]
    \centering
    \includegraphics[width=0.48\linewidth]{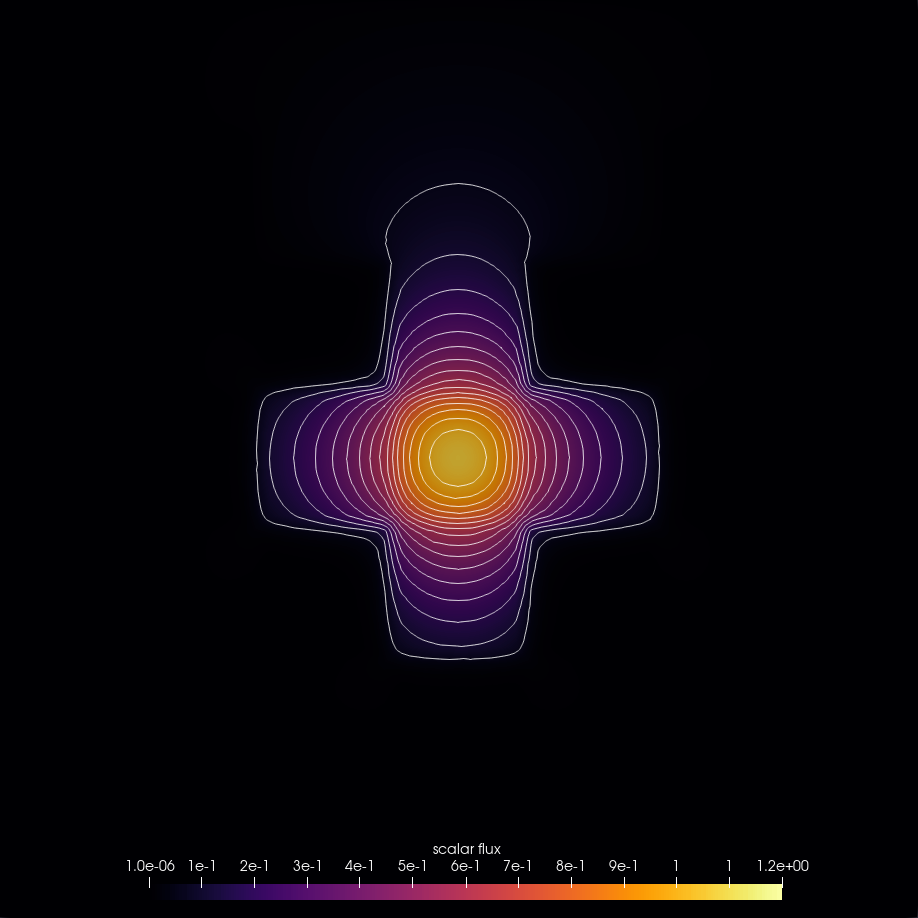}
    \includegraphics[width=0.48\linewidth]{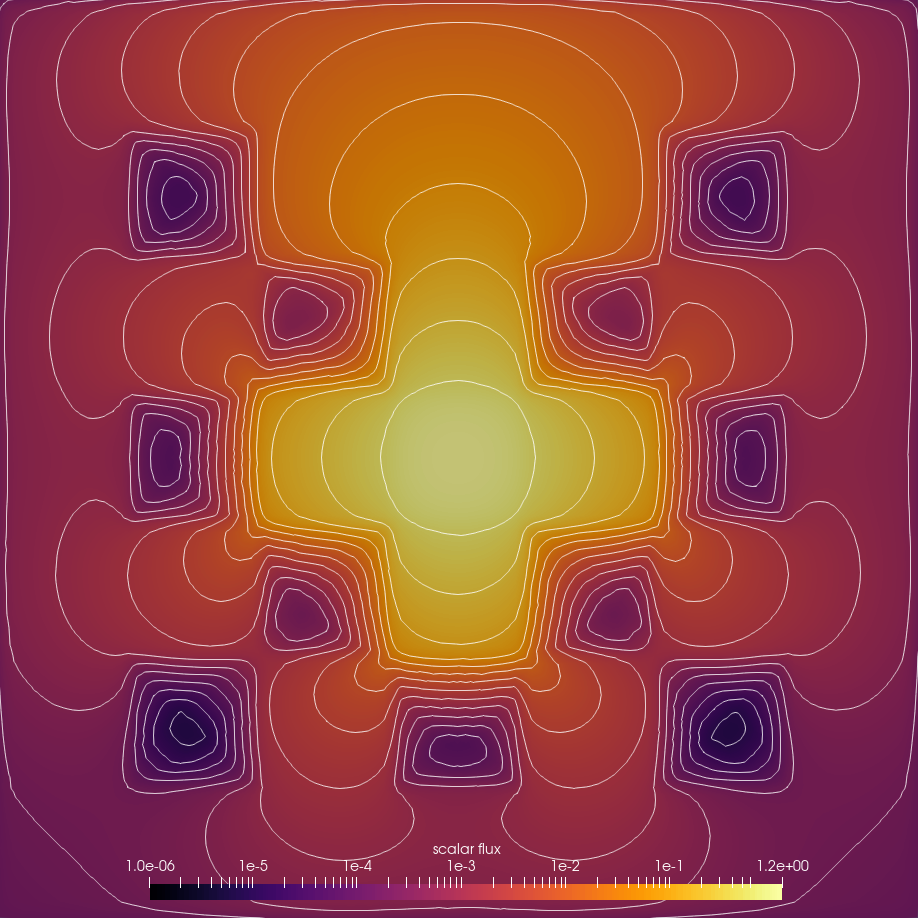}
    \caption{Simulation results for the $M_{1}$ solver with monomial basis in linear scale and log scale.}
    \label{fig_res_checker2d_M1}
\end{figure}

\begin{figure}[htp!]
    \centering
    \includegraphics[width=0.48\linewidth]{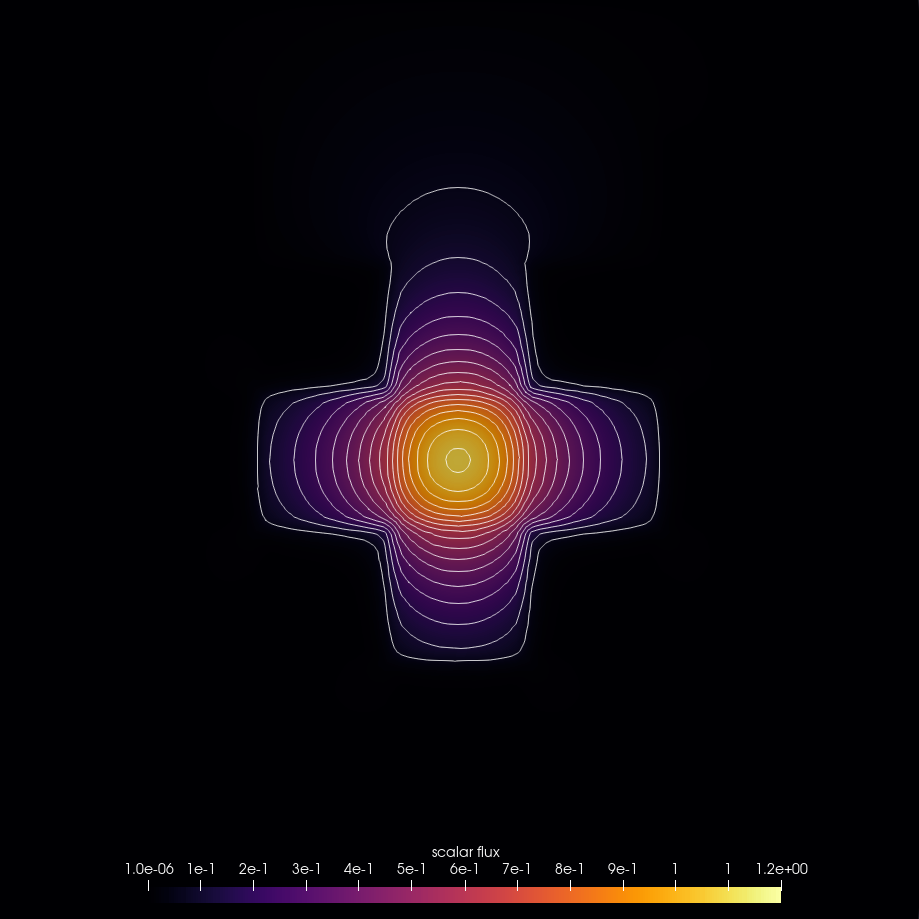}
    \includegraphics[width=0.48\linewidth]{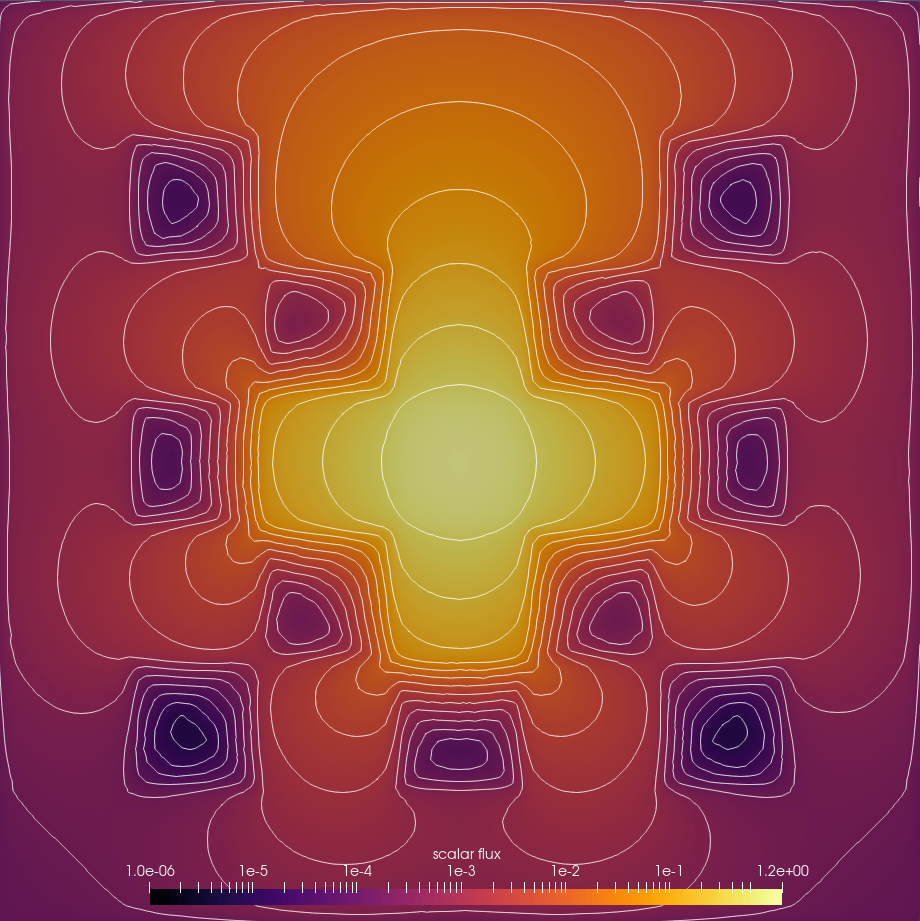}
    \caption{Simulation results for the   $M_{1}$ solver with monomial basis and neural network based entropy computation in linear scale and log scale.}
    \label{fig_res_checker2d_M1_neural}
\end{figure}

\begin{figure}[htp!]
    \centering
    \includegraphics[width=0.48\linewidth]{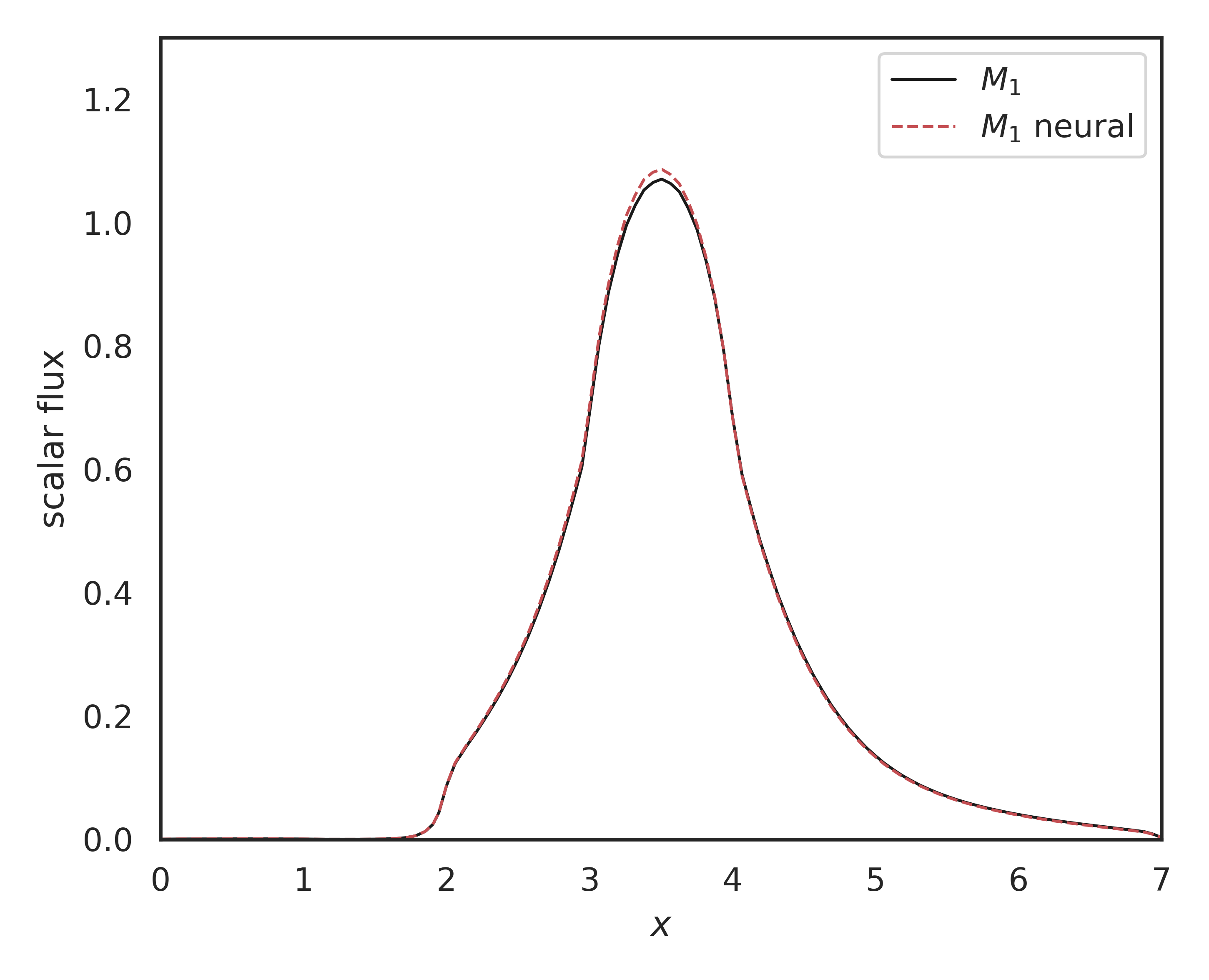}
    \includegraphics[width=0.48\linewidth]{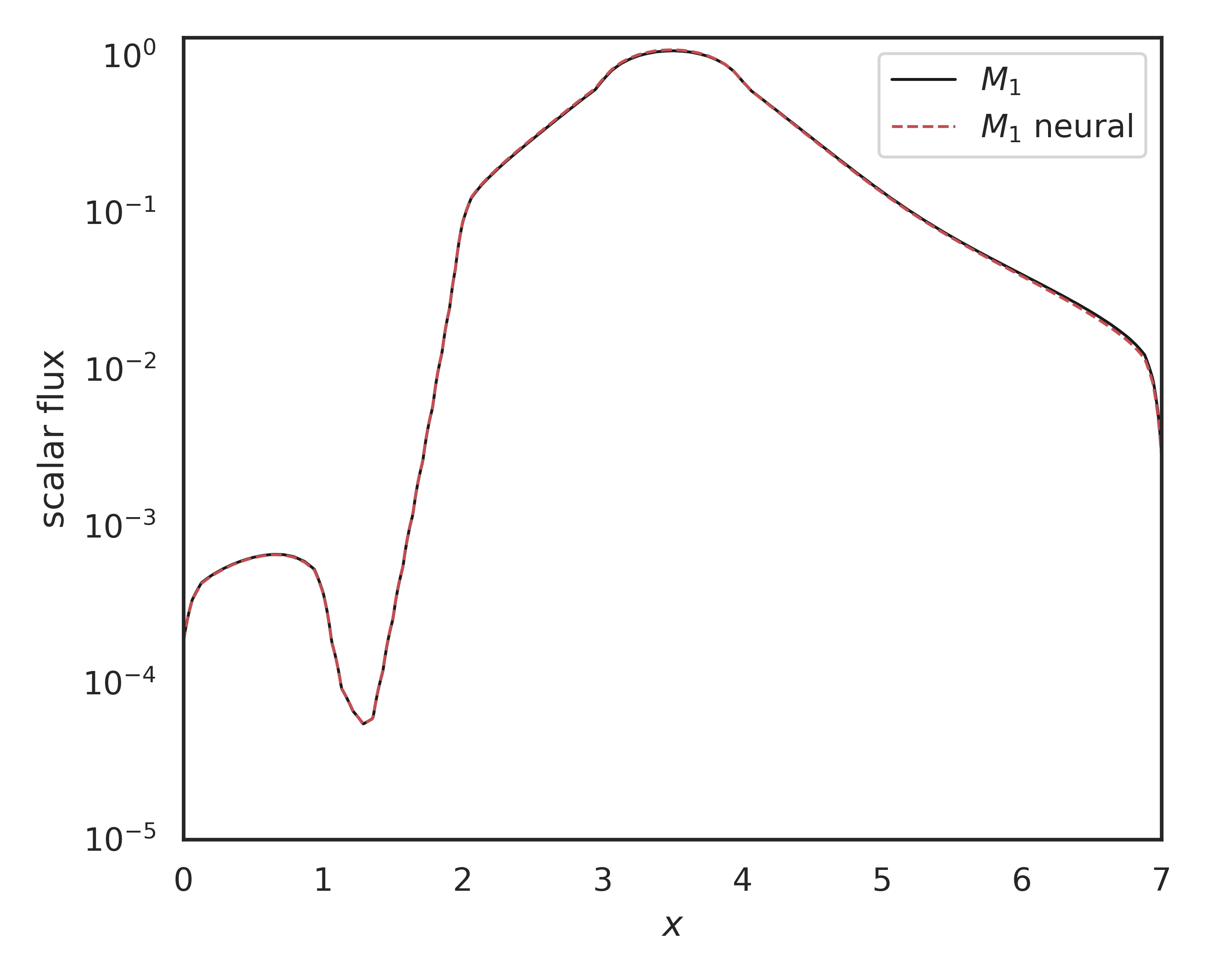}
    \caption{Cross section comparison of the Checkerboard test case using the  $M_1$ and neural network based $M_1$ method.}
    \label{fig_res_checker2d_xs_mn_neural}
\end{figure}

The checkerboard test case  mimics a nuclear reactor block with a strong radiative source in the domain center, which is denoted by $ \mathbf X_q$,  and several highly absorptive regions $ \mathbf X_a$ placed in a checkerboard pattern it. 
\begin{figure}[htp!] 
    \centering
    \includegraphics[width=0.48\linewidth]{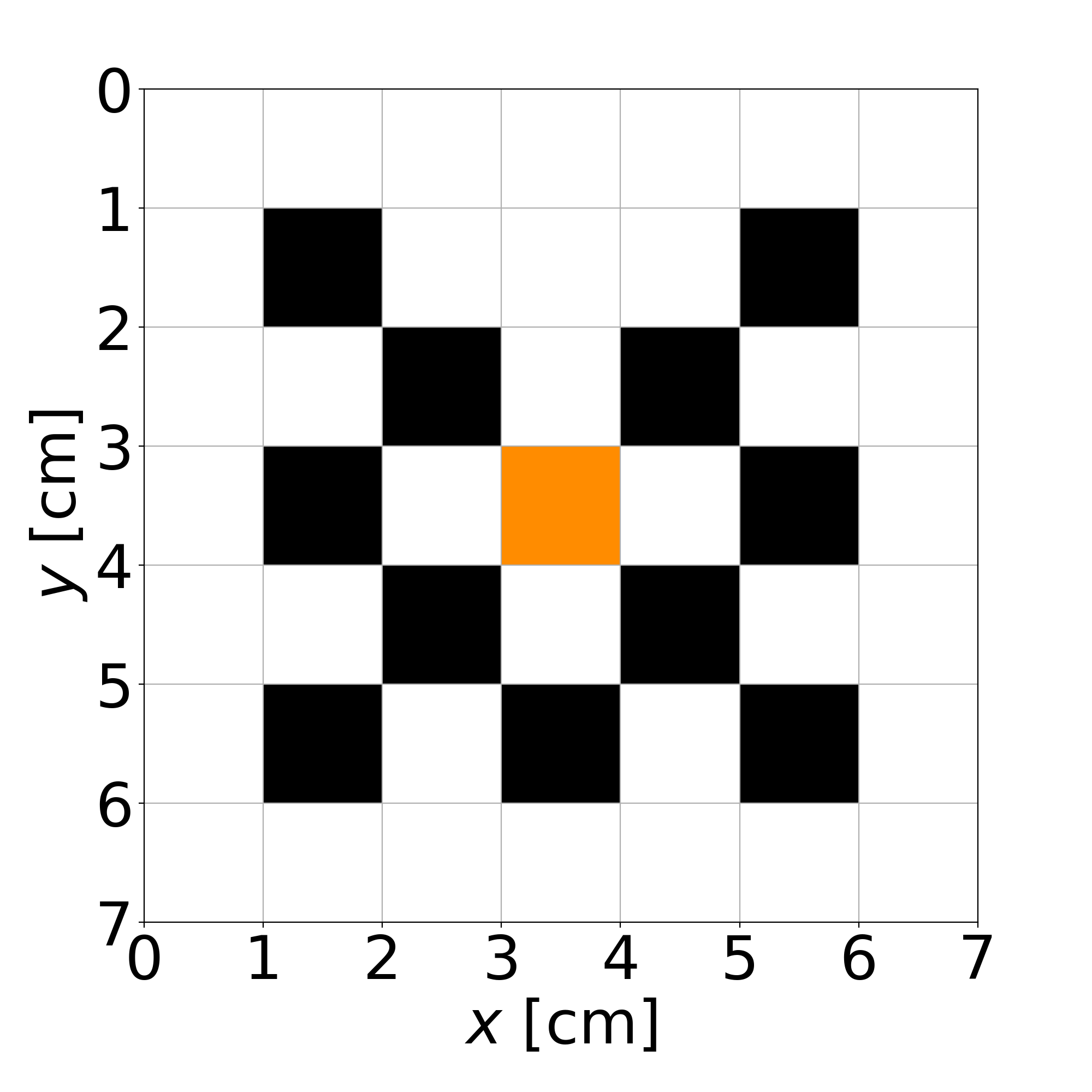}
    \caption{Layout of the two dimensional checkerboard test case.}
    \label{fig_checkerboard2d}
\end{figure}
The  spatial layout of this two dimensional test case can be found in Figure~\ref{fig_checkerboard2d}, where the source region $ \mathbf X_q$ is marked orange and the absorption regions $ \mathbf X_a$ are marked black.
The corresponding time dependent linear Boltzmann equation reads
\begin{align}
    \partial_t \psi + \mathbf{\Omega}\nabla_{ \mathbf x} \psi + \Sigma_t = \int_{\mathbb{S}^2}  \Sigma_s(t, \mathbf x,\mathbf{\Omega},\mathbf{\Omega}_*)\psi(\mathbf{\Omega}_*)\intD \mathbf{\Omega}_* + q(\mathbf{t,\mathbf x,\Omega})  
\end{align}
for  $ \mathbf x\in [0,7]^2$, $t\in[0,10)$ and $\mathbf{\Omega}\in P_{\mathbb{R}^2}(\mathbb{S}^2)$.
This corresponds to Eq.~\eqref{eq:BoltzmannCSDTrafo} with $\rho( \mathbf x)=1$.
We equip the equation with Dirichlet boundary conditions and initial condition 
\begin{align}
    \psi(t,\mathbf{x},\mathbf{\Omega})=&0,\qquad \mathbf x\in\partial \mathbf X\\
    \psi(t,\mathbf x,\mathbf{\Omega})=&0,\qquad t = 0
\end{align}
to obtain a well posed problem. Furthermore, the scattering kernel $k$ and source term $q$ are assumed to be isotropic and constant in time. The scattering cross and absorption cross sections are given by 
\begin{align}
\Sigma_s( \mathbf x) = \begin{cases}
0 & \mathbf x\in X_a\\
1 &\text{else}
\end{cases}, \qquad
\Sigma_t( \mathbf x) = \begin{cases}
10 & \mathbf  x\in  \mathbf X_a\\
1 &\text{else}
\end{cases},
\end{align}
and the isotropic source is given by
\begin{align}
q( \mathbf x, \mathbf\Omega,t) = \begin{cases}
1 & \mathbf x\in \mathbf  X_q\\
0 &\text{else}
\end{cases}.
\end{align}
We create a unstructured triangular mesh with $25000$ cells to discretize the spatial domain with regard to the absorption and source regions, such that the region boundaries coincide with the mesh faces. The simulation is computed until final time $t_f=10$ using various solver configurations.
All employed solvers use a second order upwind flux as the spatial discretization and a second order explicit Runge Kutta scheme for temporal discretization with CFL number equals $0.45$, since $M_N$ solvers with non-regularized entropy closure require a CFL number smaller than $0.5$ for stability~\cite{kusch2018maximumprinciplesatisfying,AlldredgeHauckTits, KRISTOPHERGARRETT2015573}.
The solution computed at final time $t_f=10$ is displayed in Fig.~\ref{fig_res_checker2d_SN}, where we can see the scalar flux
\begin{align}
    \Psi( \mathbf x,t) = \int_\mathbb{S}^2 \psi\intD \mathbf \Omega,
\end{align}
in the countour plot. The radiation flux is highest at the source region $ \mathbf X_q$ and almost zero in the absorption region $ \mathbf X_a$ for all solvers. Towards the top of the domain, the radiation travels  freely, whereas towards the left, right and bottom, the radiation expansion is damped by absorption regions.
Figure~\ref{fig_res_checker2d_SN} shows the solution computed with the S$_{10}$ solver with an order $10$ tensorized Gauss Legendre Quadrature. On the logarithmic scale plot, one can see ray effects right outside the chokepoints between the absorption regions, which are typical artifacts for $S_N$~\cite{Camminady2019RayEM}. We validate the S$_{10}$ solver against the implementation of the kinetic.jl framework~\cite{Xiao2021_kinetic}, and display the vertical cross section of the KiT-RT and kinetic.jl bases solution at final time $t_f$ in Fig.~\ref{fig_xs_checker_SN} using the same triangular mesh. We can see, that the deviation between implementations is below the $1\mathrm{e}{-3}$, which is the characteristic length of a grid cell.\\
\begin{figure}[htp!]
    \centering
    \includegraphics[width=0.48\linewidth]{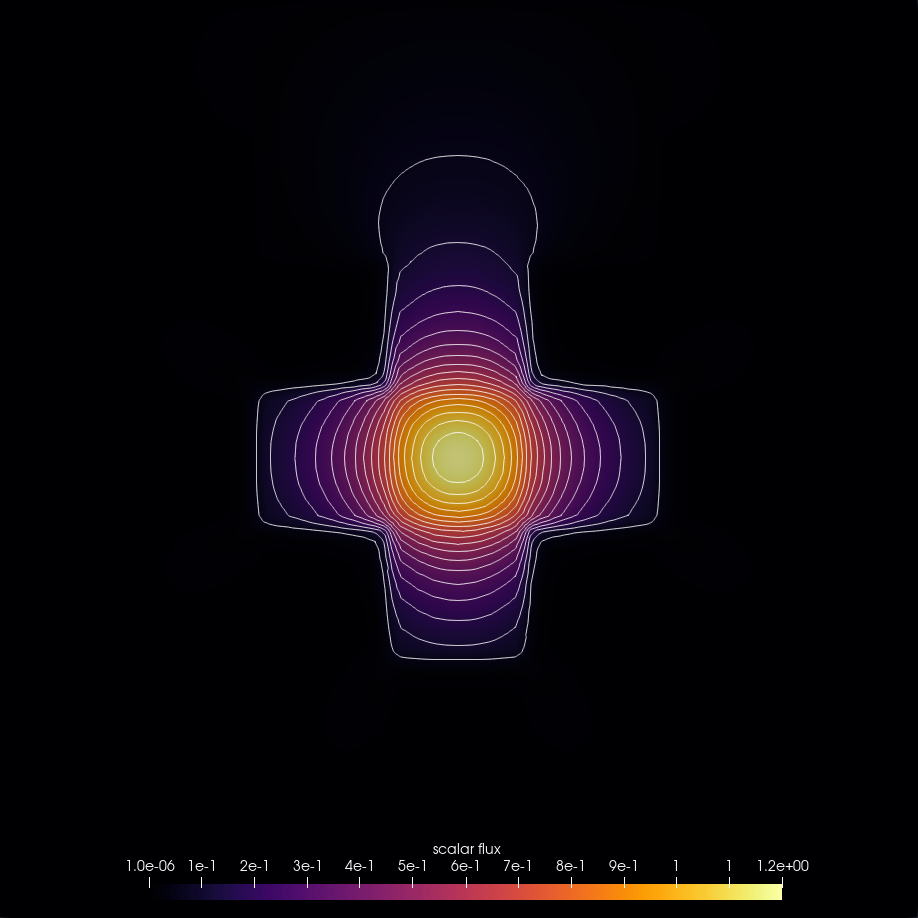}
    \includegraphics[width=0.48\linewidth]{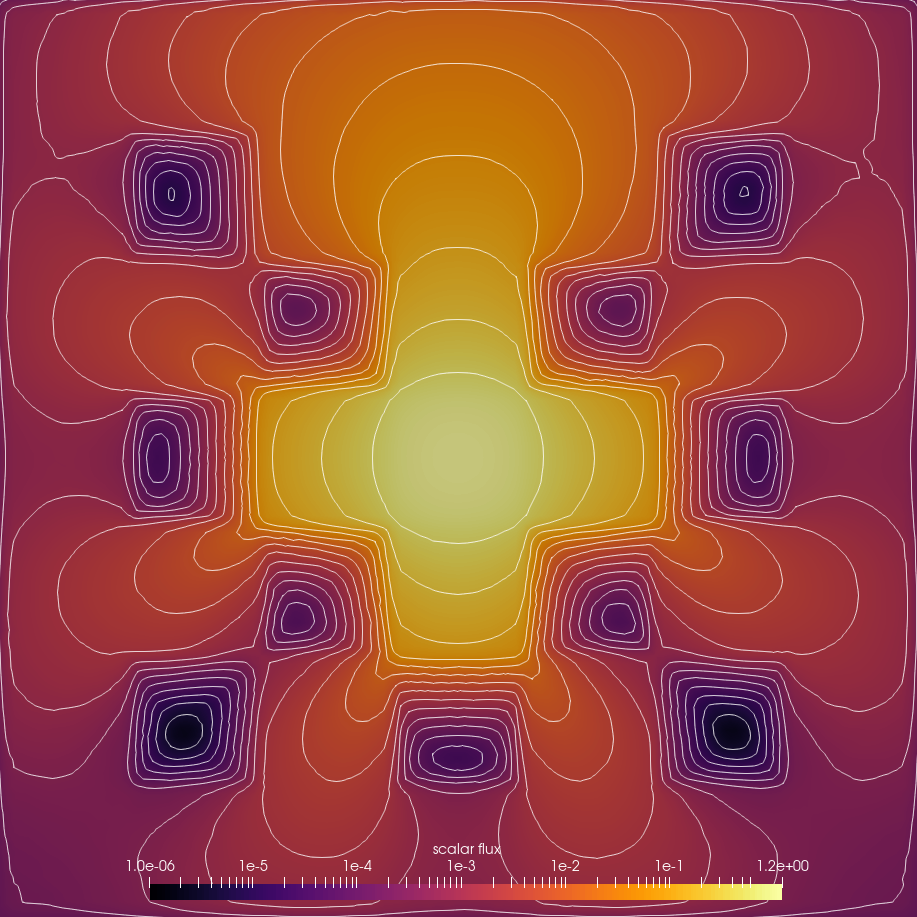}
    \caption{Simulation results for the P$_{5}$ solver in linear scale and log scale.}
    \label{fig_res_checker2d_PN}
\end{figure}
Figure~\ref{fig_res_checker2d_PN} displays the solution computed with the $P_5$ solver using an order $5$ Spherical Harmonics basis. Compared to the $S_{10}$ solution, there are no ray effects visible at the chokepoints between absorption regions.\\
\begin{figure}[htp!]
    \centering
    \includegraphics[width=0.48\linewidth]{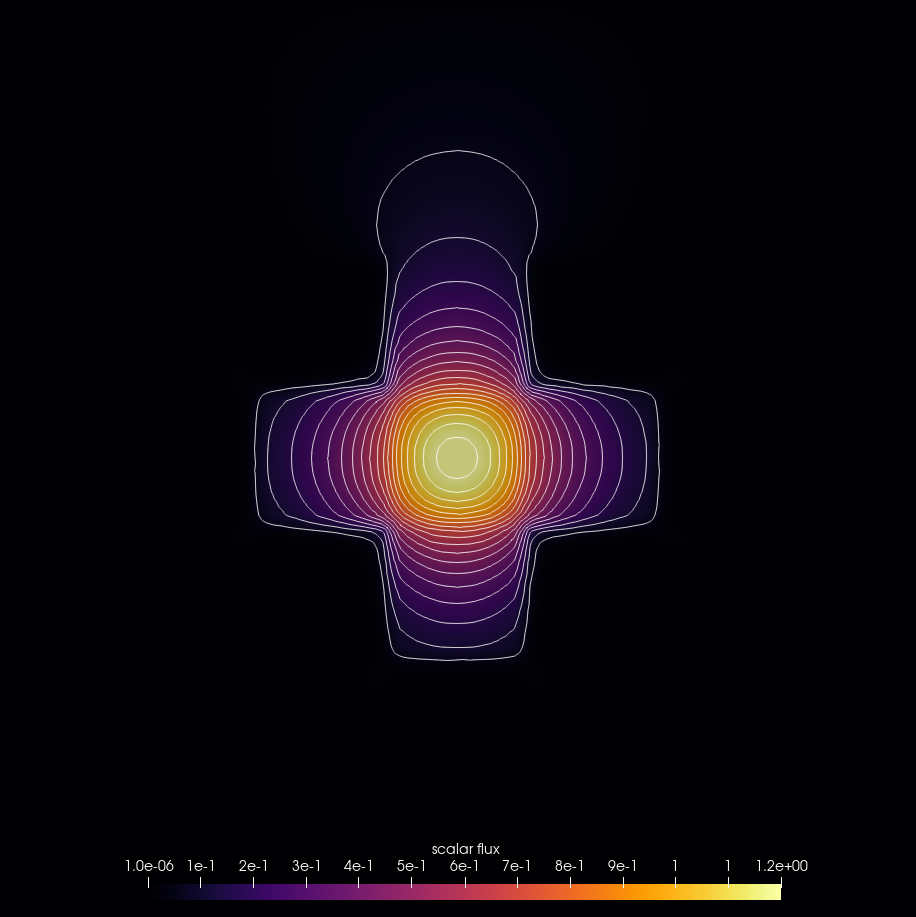}
    \includegraphics[width=0.48\linewidth]{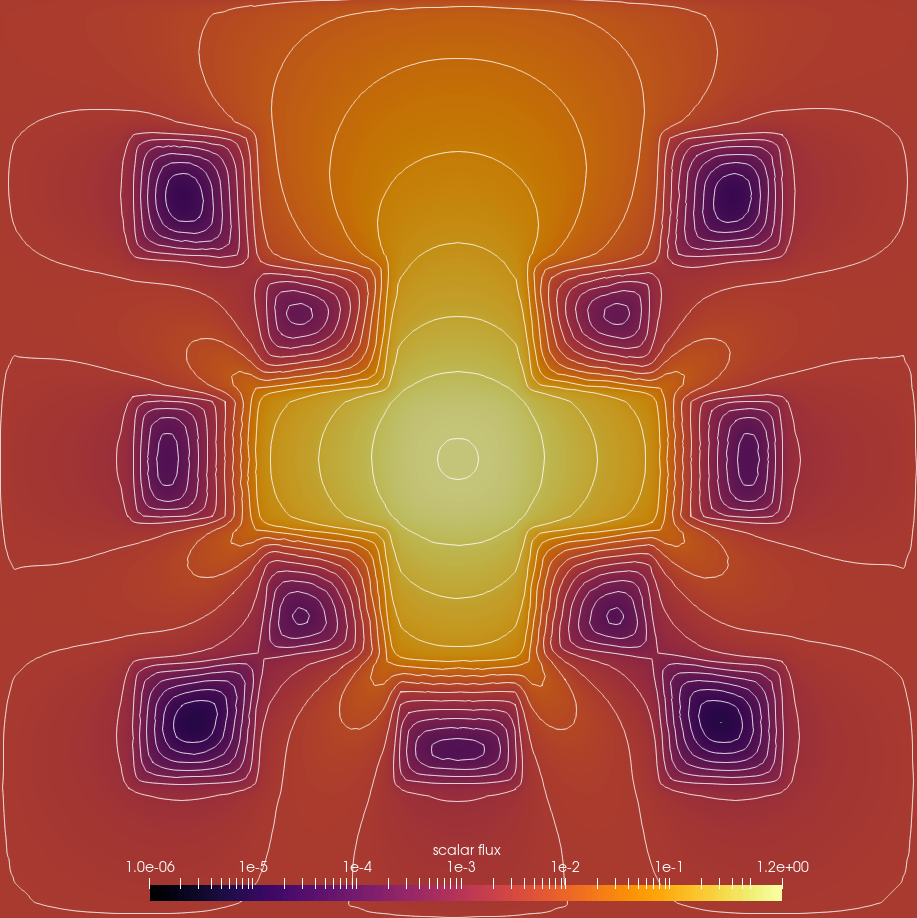}
    \caption{Simulation results for the M$_{3}$ solver with spherical harmonics basis in linear scale and log scale.}
    \label{fig_res_checker2d_MN}
\end{figure}
\begin{figure}[htp!]
    \centering
    \includegraphics[width=0.48\linewidth]{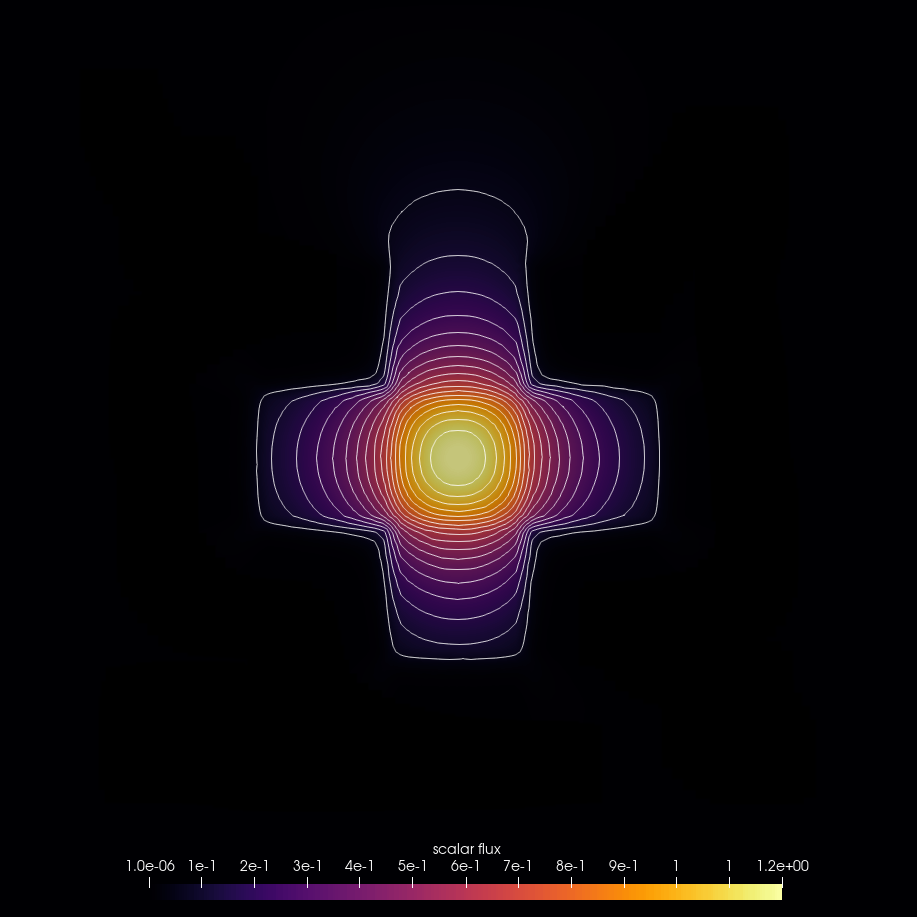}
    \includegraphics[width=0.48\linewidth]{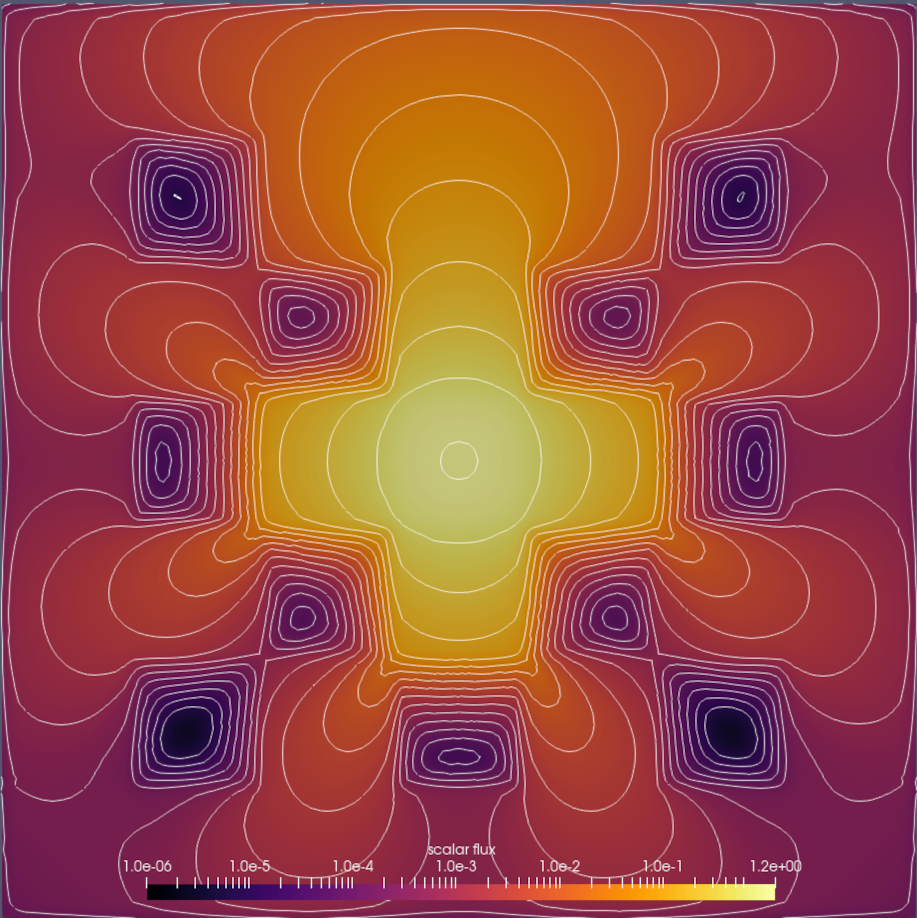}
    \caption{Simulation results for the  M$_{3}$ solver with spherical harmonics basis and regularized entropy in linear scale and log scale.}
    \label{fig_res_checker2d_MN_reg}
\end{figure}
We show now the solution for different entropy based moment, i.e., $M_N$  solvers, that use an order $10$ tensorized Gauss Legendre quadrature to evaluate the kinetic flux. Figure~\ref{fig_res_checker2d_MN} displays a M$_3$ solution using an order $3$ spherical harmonics moment basis computed with a Newton optimizer with line-search configured to accuracy $1\mathrm{e}{-7}$.
The same solver configuration is displayed in Figure~\ref{fig_res_checker2d_MN_reg}, where the entropy closure is replaced by the partial regularized minimal entropy problem of 
Eq.~\eqref{eq_entropyOCP_part_reg} with the  regularizer $\gamma$ set to $1e-3$ and a non-regularized moment of order zero. While on the linear scale plots, both solutions are similar, on the log scale plots, one can see that the regularized $M_N$ solver computes much lower values within the absorption regions.  Furthermore, the log plot shows that the regularized $M_N$ solution oscillates slightly in regions with small scalar fluxes. Figure~\ref{fig_xs_checker_solvers} shows a direct comparison of a vertical cross section using all four presented solvers. In the vast majority of the spatial domain, the solutions correspond well, with the exception of absorption regions, i.e., $x\in[1,2]$, where the $M_3$ solution computes slightly bigger values and the regularized $M_3$ solution captures the absorption region slightly better than the reminding solvers. \\
Next we compare a non-regularized M$_1$ solver using a monomial basis of order $1$ and a Newton solver based entropy computation, shown in Fig.~\ref{fig_res_checker2d_M1},  with a similar solver configuration, which employs a neural network to compute the entropy closure, shown in Fig~\ref{fig_res_checker2d_M1_neural}. Various data driven entropy closures have been introduced  by~\cite{schotthoefer2021structurepreserving,porteous2021datadriven}. In the KiT-RT package we have compiled the trained networks by~\cite{schotthoefer2021structurepreserving} to C++ and have implement a fast tensorflow~\cite{tensorflow2015-whitepaper} backend for seamless integration into the KiT-RT framework. The neural network used in the simulation of Fig~\ref{fig_res_checker2d_M1_neural} is an input-convex architecture. Both, the Newton based and neural network based solution correspond well. As Fig.~\ref{fig_res_checker2d_xs_mn_neural} shows, only in at the top of the domain, i.e., at $x\approx 6.5$ in the cross section plot, we see a small deviation between the methods. In this region at the wave-front of the radiative transport, the moments of the kinetic equation are close to the boundary of the realizable set, where on the one hand, the Newton based solver needs more iterations and thus more wall time to compute the solution to the optimization problem, and on the other hand, the neural network accuracy declines. \\
Neural network based entropy closures are constructed to accelerate the time consuming $M_N$ method. In the following we validate the speedup through the neural network entropy closure using a  larger mesh of $700000$ grid cells, since in ~\cite{schotthoefer2021structurepreserving} it is shown, that the due to data transformation to tensorflow tensors, the speedup is faster for higher data-set sizes. The time consumption of the $M_1$ and neural network based $M_1$ solver is illustrated in Table~\ref{tab_timingbenchmark}, where one can see that the neural network based closure accelerates the compuational time by $89.33$-$87.01$\%.
\begin{table}[htbp]
	\caption{Timings of the neural network based $M_1$ solver compared to the Newton based $M_1$ solver}
	\centering
	\begin{tabular}{|l|l|l|l|} 
		\hline
		 & $M_1$ Newton & $M_1$ neural network & runtime reduction   \\
		\hline 
	    $4$ cores & $757.88979$  & $80.810305$ & $89.33$\% \\ 
	    \hline
	    $12$ cores & $258.6485$  & $33.606152$ & $87.01$\% \\ 
		\hline
	\end{tabular} 
	\label{tab_timingbenchmark}
\end{table}

\subsection{Beam in 2D patient CT}

Having validated the CSD solvers against StarMAP and a Monte Carlo framework in section \ref{sec:inhom_linesource}, we now examine a realistic 2D CT scan of a lung patient as a proof of concept for the application of our framework to radiation therapy computations. The patient data was retrieved from an open source data set \cite{wang2020data} in The Cancer Imaging Archive (TCIA) \cite{clark2013archive}. The patient is irradiated with an electron beam of $E_{\text{max}} = 20$ MeV. We model this beam as the initial condition
\begin{align*}
    \psi(E_{max},\mathbf x, \mathbf \Omega) = \frac{1}{(2\pi)^{3/2}\sigma_{\Omega_2}\sigma_x\sigma_y} \cdot\exp(-(\mu_{\Omega_2}-\Omega_2)^2/2\sigma_{\Omega_2})
    \cdot\exp(-(\mu_{x_1}-x_1)^2/2\sigma_{x_1})\cdot\exp(-(\mu_{x_2}-x_2)^2/2\sigma_{x_2}),
\end{align*}
where $(\mu_{x_1},\mu_{x_2})=(2.5\text{cm},5.8\text{cm})$ is the beam position within the $6\text{cm}x6\text{cm}$ domain and $\mu_{\Omega_2}=\frac{\pi}{2}\text{rad}$ is the beam direction. The remaining parameters are chosen as $\sigma_{x_1}=\sigma_{x_2}=\sigma_{\Omega_2}=0.1$. To determine a tissue density $\rho$ for given gray-scale values of the CT image, we set the maximum density, represented by white pixels, to the density of bone $\rho_{\text{bone}} = 1.85\text{ g}/\text{cm}^3$. The remaining tissue is scaled such that the minimum pixel value of zero corresponds to a minimal density of $\rho_{\text{min}} = 0.05 \text{ g}/\text{cm}^3$. This corresponds approximately to the lower bound of observed lung densities \cite{kohda1989measurement}.

\begin{figure}[htp!]
\centering
\begin{subfigure}{.48\textwidth}
\centering
  \includegraphics[width=.95\linewidth]{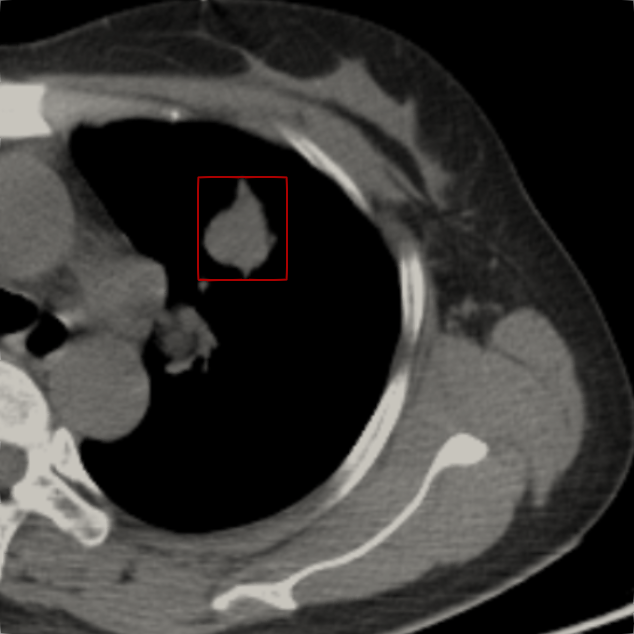}
  \caption{ CT scan}
  \label{fig:sub1}
\end{subfigure}%
\begin{subfigure}{.48\textwidth}
\centering
  \includegraphics[width=.95\linewidth]{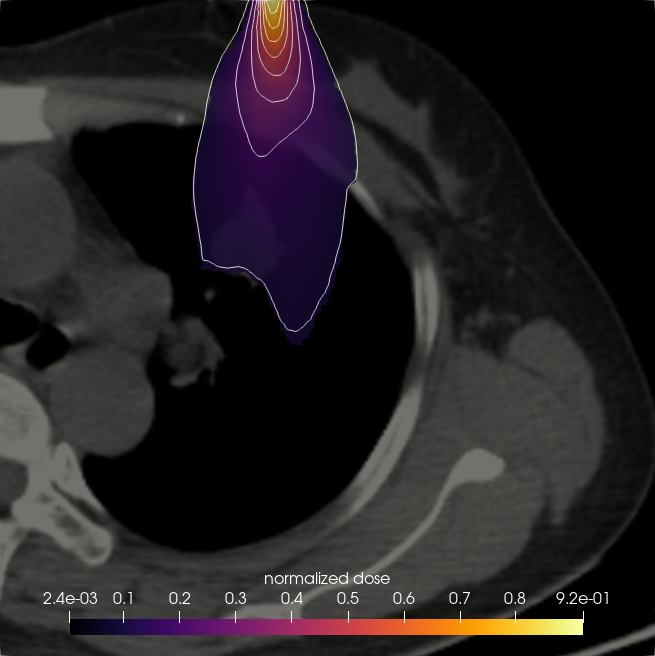}
  \caption{ $S_{13}$}
  \label{fig:sub2}
\end{subfigure}
\begin{subfigure}{.48\textwidth}
\centering
  \includegraphics[width=.95\linewidth]{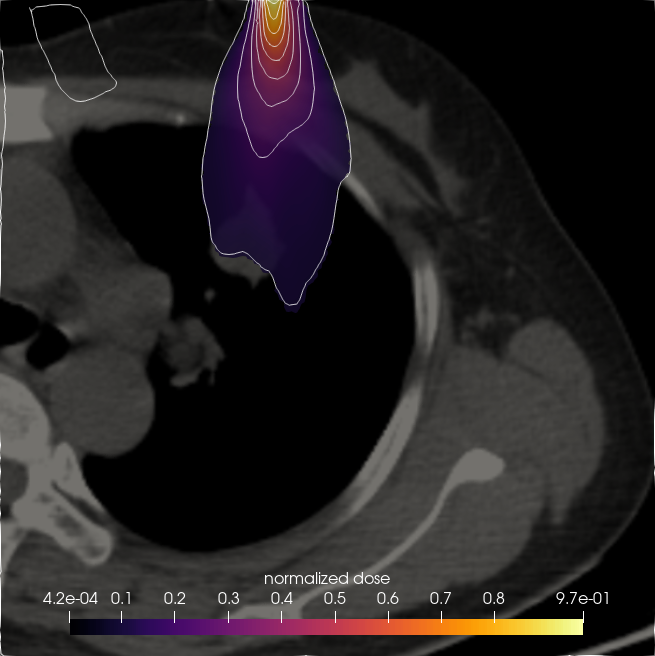}
  \caption{$P_{13}$ }
  \label{fig:sub3}
\end{subfigure}%
\begin{subfigure}{.48\textwidth}
\centering
  \includegraphics[width=.95\linewidth]{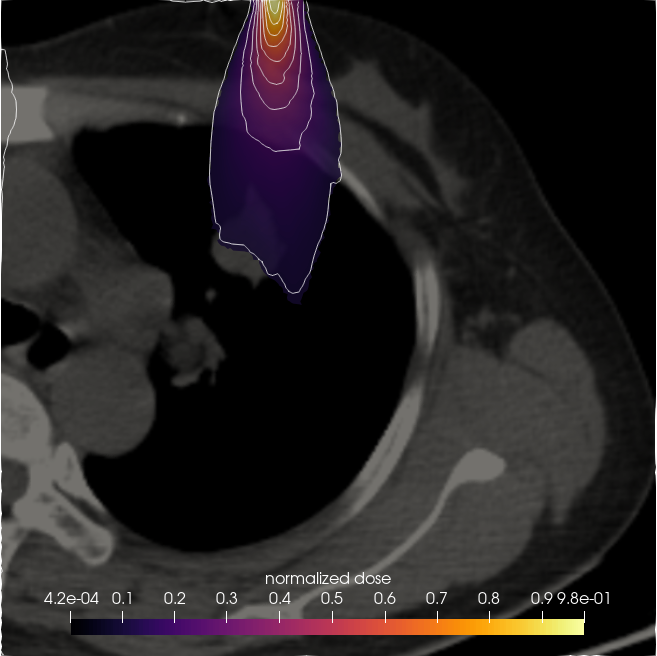}
  \caption{$M_{5}$}
  \label{fig:sub4}
\end{subfigure}
\caption{Patient CT scan with lung tumor (red box) (a) as well as corresponding simulation results for the $S_{13}$, $P_{13}$ and $M_{5}$ solver with spherical harmonics basis and partially regularized entropy.}
\label{fig:lungCT}
\end{figure}

\begin{figure}[htp!]
    \centering
    \includegraphics[width=0.48\linewidth]{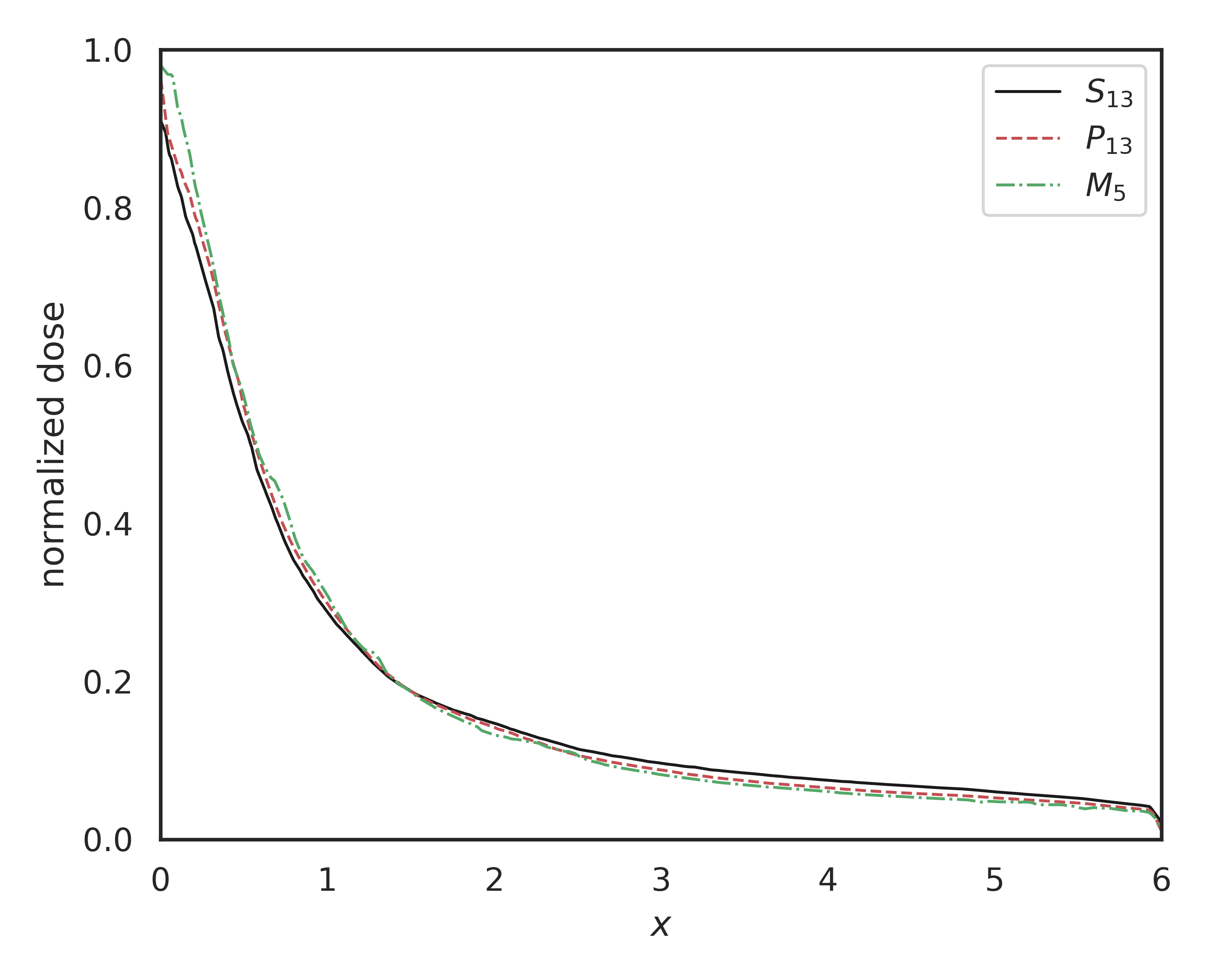}
    \includegraphics[width=0.48\linewidth]{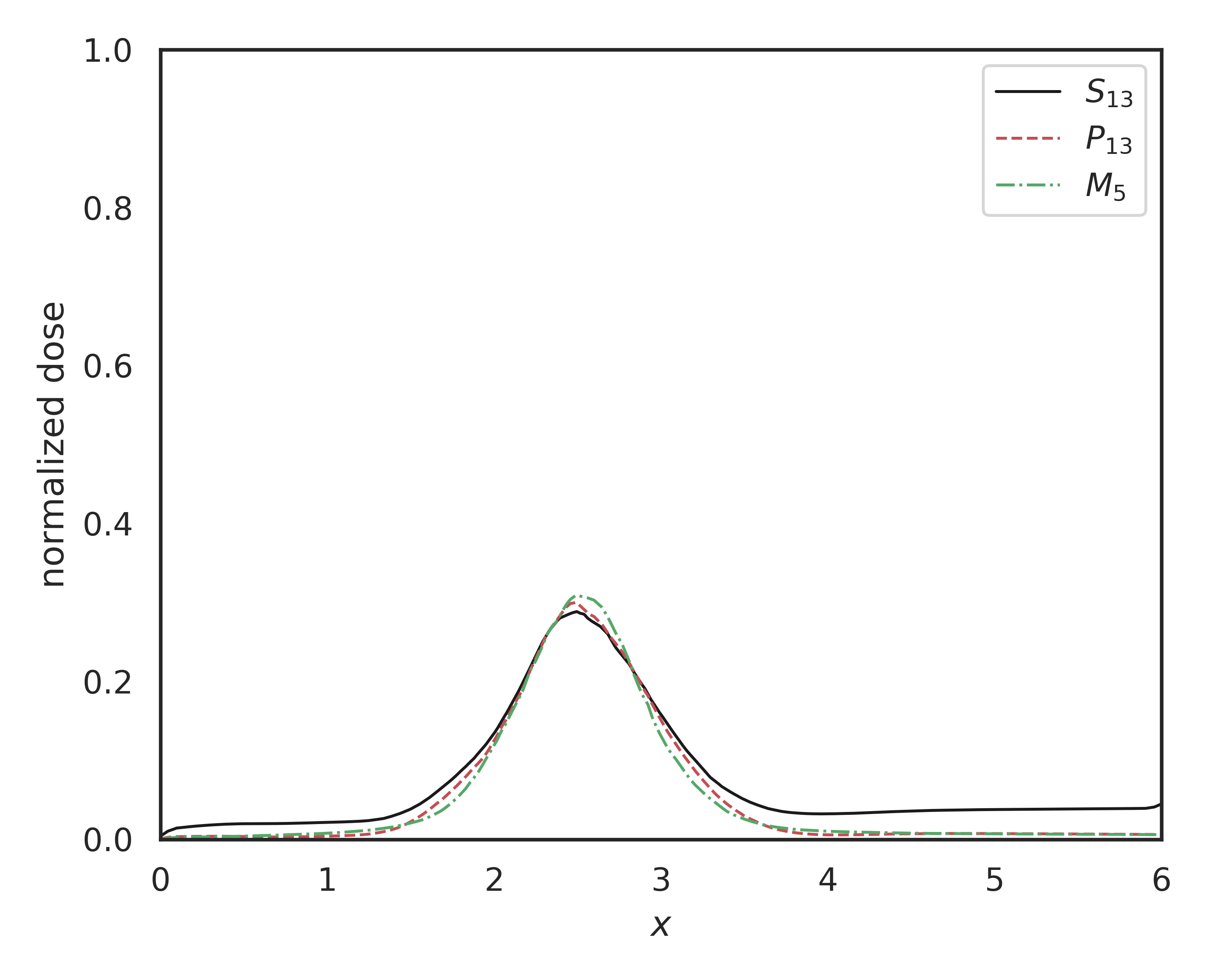}
    \caption{Vertical (at $x=2.5$cm) and horizontal (at $y = 5$cm) cross section through the normalized dose in the patient CT. Comparison of the  $S_{13}$, $P_{13}$ and partially regularized $M_{5}$ solver.}
    \label{fig:xs_lungCT_solvers}
\end{figure}

Figure \ref{fig:lungCT} compares the normalized dose for a CSD $S_{13}$, $P_{13}$ and $M_5$ solver. While all methods show similar behaviour and are able to capture the effects of heterogeneities in the patient density, some differences e.g. in the maximum depth of the $S_{13}$ solution compared to $P_{13}$ and $M_5$ or the shape of the  lowest two isolines can be observed. The cross sections in figure \ref{fig:xs_lungCT_solvers} further show that the $S_{13}$ dose has a lower maximum and higher minimum value than the $M_5$ and to a lesser extent also $P_{13}$ solutions. 

\section{Conclusion}
In this work, we have presented a collection of deterministic transport solvers for radiation therapy applications. The methods agree well with results obtained with conventional radiation therapy codes. Due to the use of polymorphism, we are able to guarantee a straight forward extension to further numerical methods, which facilitates the investigation of novel radiation therapy solvers and their comparison to conventional methods.
\section*{Acknowledgements}
All authors of this work have contributed equally to this project.
The authors would like to thank Thomas Camminady for his help with implementing spherical quadrature rules. Jonas Kusch has been funded by the Deutsche Forschungsgemeinschaft (DFG, German
Research Foundation) – 491976834. Pia Stammer is supported by the Helmholtz Association under the joint research school HIDSS4Health -- Helmholtz Information and Data Science School for Health.
The work of Steffen Schotthöfer is funded by the Priority Programme "Theoretical Foundations of Deep Learning (SPP2298)"  by the Deutsche Forschungsgemeinschaft.

\bibliographystyle{ACM-Reference-Format}
\bibliography{ref}

\end{document}